\documentclass[a4paper]{JHEP3}
\pdfoutput=1
\usepackage{amsmath,amssymb,graphicx}
\usepackage{epsf}
\usepackage{subfigure}
\usepackage{cite}
\graphicspath{{./Figs/}{./Figs_OLD/}}

\newcommand{\nn}{\nonumber}
\newcommand{\beq}{\begin{equation}}
\newcommand{\eeq}{\end{equation}}
\newcommand{\bea}{\begin{eqnarray}}
\newcommand{\eea}{\end{eqnarray}}
\newcommand{\<}{\langle}
\renewcommand{\>}{\rangle}
\newcommand{\ov}{\overline}
\def\sla#1{\setbox0=\hbox{$#1$}\dimen0=\wd0
      \setbox1=\hbox{/} \dimen1=\wd1 \ifdim\dimen0>\dimen1
      \rlap{\hbox to \dimen0{\hfil/\hfil}} #1                        \else
      \rlap{\hbox to \dimen1{\hfil$#1$\hfil}}
      /   \fi}
\newcommand{\eps}{\epsilon}
\newcommand{\te}{\theta}

\newcommand{\Tr}{{\rm Tr}}
\newcommand{\id}{1 \hspace{1.15mm} \!\!\!\!1}
\newcommand{\mc}{\mathcal}
\newcommand{\re}{{\rm Re}}
\newcommand{\im}{{\rm Im}}

\preprint{LPT Orsay 11-22\\UM-DOE/ER/40762-493}

\title{Gauged Flavor Group with Left-Right Symmetry}

\author{Diego Guadagnoli$^a$, Rabindra~N.~Mohapatra$^b$ and Ilmo Sung$^b$\\
$^a$ Laboratoire de Physique Th\'eorique, Universite Paris-Sud, Centre d'Orsay, F-91405 Orsay-Cedex, France\\
$^b$ Maryland Center for Fundamental Physics, Department of Physics, University of Maryland, College Park, MD 20742, USA\\
Email: \email{diego.guadagnoli@th.u-psud.fr, rmohapat@umd.edu, sung@umd.edu}\\
(Dated: \today)}

\abstract{We construct an anomaly-free extension of the left-right
symmetric model, where the maximal flavor group is gauged and
anomaly cancellation is guaranteed by adding new vectorlike
fermion states. We address the question of the lowest allowed
flavor symmetry scale consistent with data. Because of the
mechanism recently pointed out by Grinstein {\em et al.}
tree-level flavor changing neutral currents turn out to play a very
weak constraining role. The same occurs, in our model, for electroweak
precision observables. The main constraint turns out to come from
$W_R$-mediated flavor changing neutral current box diagrams,
primarily $K - \ov K$ mixing. 
In the case where discrete parity symmetry is present 
at the TeV scale, this constraint implies lower bounds on the mass
of vectorlike fermions and flavor bosons of 5 and 10 TeV
respectively. However, these limits are weakened under the
condition that only $SU(2)_R\times U(1)_{B-L}$ is restored at the
TeV scale, but not parity. For example, assuming the $SU(2)$
gauge couplings in the ratio $g_R / g_L \approx 0.7$ allows the
above limits to go down by half for both vectorlike fermions
and flavor bosons. Our model provides a framework for
accommodating neutrino masses and, in the parity symmetric case,
provides a solution to the strong CP problem. The bound on the
lepton flavor gauging scale is somewhat stronger, because of Big
Bang Nucleosynthesis constraints. We argue, however, that the
applicability of these constraints depends on the mechanism at
work for the generation of neutrino masses.}

\begin{document}

\section{Introduction} \label{sec:intro}

One of the long-standing mysteries of physics beyond the Standard
Model is the origin of flavor patterns for quarks and leptons. In
the Standard Model~(SM), they arise from the quark and lepton
Yukawa couplings with the SM Higgs boson and are arbitrary,
thereby precluding any physical insight as to their origin. Since
these flavor patterns may well be the remnants of the breaking of
some symmetry, the belief is that pinning down the flavor symmetry
at work could provide hints of the underlying dynamics at work.
Many possibilities for approaching this important issue from the
vantage point of symmetry then present themselves -- starting from
discrete non-abelian subgroups of these flavor symmetries to
continuous global or local ones. The question then arises as to
how we determine by low energy observations which particular
mechanism is at work and at which scale such a symmetry manifests
itself. The hope is that different choices will lead to different
characteristic predictions, e.g.\ a global horizontal symmetry
would lead to massless familons at low energies \cite{wilczek} and
discrete symmetries could lead to some relations between
observables.

A widely discussed possibility is to study gauged flavor
symmetries \cite{gauged}, which leads to a number of interesting
effects such as new gauge bosons and new flavor changing effects
mediated by these bosons. The very first test of this possibility
is to determine the scale of gauged flavor symmetry. Na\"{i}ve
considerations seem to suggest that this scale is likely to be in
the 1000 TeV range; however in specific models this expectation
could change drastically. For example, in a recent paper by
Grinstein, Redi and Villadoro (GRV) \cite{GRV}, it has been shown
by explicit construction that there are SM extensions with
gauged flavor symmetry where this scale could be in the 1 TeV
range or even below, compatible with constraints from the hadronic
sector. Furthermore this model predicts, by the requirement of
anomaly cancellation, new vectorlike quarks,
the lightest of which with masses again in the TeV ballpark, hence
within the reach of LHC direct searches. The mechanism at work in
the GRV model is an inverse, see-saw like, relation
between the masses of the quarks and those of the new fermion
states \cite{babu} as well as of the flavor gauge bosons, so that
the partners of the heaviest quarks are the lightest among
the new states. 
This fact allows to pass 
all the flavor changing neutral current~(FCNC) constraints in a very natural way.

A certain degree of model dependence in the idea of gauging the
flavor symmetry is the choice of new fermionic states added in
order to cancel the gauge anomalies. To achieve this, the simplest
option is to include new, vectorlike quark partners. This choice
may however appear to be at odds with that of quark states that
need sit in chiral representations. This simple consideration motivates
us to pursue here an alternative possibility, where the maximal
gauging of flavor symmetry is carried out within a left-right
symmetric extension of the SM \cite{LR}. This implies that the flavor
symmetry group is itself left-right symmetric. The new gauge
anomalies resulting from the larger gauge group cancel with the
introduction of vectorlike new fermionic partners of
the quarks \cite{babu}. Besides the known virtues inherent in
left-right symmetric extensions of the SM, e.g. the possibility to
justify the hypercharge quantum numbers, there appear to be
the following advantages in our approach: {\em (i)} it provides a
natural way to include neutrino masses; {\em (ii)} in the quark
Lagrangian, it features only three free parameters in the gauge and
Yukawa sector, after the rest are fixed by data on quark masses
and mixings; {\em (iii)} the model provides a simple solution to
the strong CP problem without the need for an axion, in a manner
similar to that discussed in Refs.\ \cite{ms,babu1}.

There are two possible realizations of this idea while keeping the
gauge group to be $SU(2)_L\times SU(2)_R\times U(1)_{B-L}$ at the
TeV scale:
\textit{(a)} the discrete parity symmetry is maintained down to the
TeV scale or
\textit{(b)} it is broken at some very high scale \cite{CMP} so that, at the
TeV scale, the two gauge couplings as well as the left and right Yukawa
couplings are in general different from each other. We will consider
both alternatives below.
Within the second alternative, the solution to the strong CP problem
mentioned above is not obvious.

Concerning the constraints on the model outlined above, we find that,
similarly as in the GRV model, tree-level FCNCs mediated by the flavor
gauge bosons are tamed automatically by the hierarchy of their masses.
Also, in our case, electroweak (EW) precision tests are automatically fulfilled
in the bulk of the parameter space.

The strongest constraint comes from $W_R$-mediated FCNC box diagrams, 
primarily $K - \ov K$ mixing. The implied bounds on the scale of the new 
vectorlike quarks as well as on the flavor gauge bosons, which will be 
discussed in detail in secs.\ \ref{sec:WRmass}, \ref{sec:gaugemassscale} 
and \ref{sec:fermionmixing}, depend however on the scale at which parity
is broken.

Compatibly with these bounds, the new effects, accessible at
the current generation of TeV hadron colliders, include the
lightest among the new particles' masses and various deviations in
top-physics observables, like top production and decays. Such
deviations are due to the fact that the top (in particular, the
right-handed one) mixes non-negligibly onto its new fermionic
partner, whereas mixing is tiny to absent for the rest of the
quark states. This in turn explains why no deviations are to be
expected in the production or decays of any other quark than
$t_R$.

For an overview of the organization of this paper we refer the reader
to the table of content on page\ \pageref{sec:intro}.

\section{Flavor Symmetry within Left-Right Models} \label{sec:model}

In the SM, once the Yukawa couplings are set to zero,
the maximal flavor symmetry group is $SU(3)_{Q_L} \times
SU(3)_{u_R}\times SU(3)_{d_R}\times SU(3)_{\ell_L}\times
SU(3)_{\ell_R}$. If the weak gauge group is extended
to that of the left-right symmetric model, the flavor group becomes
$SU(3)_{Q_L} \times SU(3)_{Q_R}\times SU(3)_{\ell_L}\times
SU(3)_{\ell_R}$ which is more economical and, unlike the SM, also
simultaneously explains neutrino masses.\footnote{For other
horizontal symmetry extensions of left-right models, see
\cite{references}. Furthermore, gauged flavor symmetries
have also been discussed in the context of the Pati-Salam GUT
in ref. \cite{feldi}, see appendix therein.}

We will therefore start with the gauge group $G_{LR} \equiv
SU(3)_{c} \times SU(2)_L \times SU(2)_R \times U(1)_{B-L} \times
SU(3)_{Q_L} \times SU(3)_{Q_R}\times SU(3)_{\ell_L}\times
SU(3)_{\ell_R}$, where $SU(3)_{Q_L} \times SU(3)_{Q_R}$
represents the flavor gauge symmetries respectively in
the left- and right-handed quark sector, and $SU(3)_{\ell_L}
\times SU(3)_{\ell_R}$ the corresponding ones for the lepton sector.
The particle content and its transformation properties under
fundamental representations of the group $G_{LR}$ are reported in
table \ref{tab:model}.
\begin{table}
\small
\begin{center} \begin{tabular}{ l | c  c c c  c c c c }
  & $SU(2)_L$ & $SU(2)_R$ & $U(1)_{B-L}$ & $SU(3)_{Q_L}$ &
  $SU(3)_{Q_R}$ & $SU(3)_c$ & $SU(3)_{\ell_L}$ & $SU(3)_{\ell_R}$ \\ [3pt]
\hline \hline
  $Q_L$ & 2 & & $\frac{1}{3}$ & 3 &   & 3 & & \\ [3pt]
  $Q_R$ & & 2 & $\frac{1}{3}$ &   & 3 & 3 & & \\ [3pt]
  $\psi^{u}_{L}$ & & & $\frac{4}{3}$   &   & 3 & 3 & & \\ [3pt]
  $\psi^{u}_{R}$ & & & $\frac{4}{3}$   & 3 &   & 3 & & \\ [3pt]
  $\psi^{d}_{L}$ & & & $-\frac{2}{3}$  &   & 3 & 3 & & \\ [3pt]
  $\psi^{d}_{R}$ & & & $-\frac{2}{3}$  & 3 &   & 3 & & \\ [3pt]
\hline
  $L_L$ & 2 & & $-1$ & & & & 3 &   \\ [3pt]
  $L_R$ & & 2 & $-1$ & & & &   & 3 \\ [3pt]
  $\psi^{e}_{L}$ & &   & $-2$ & & & &   & 3 \\ [3pt]
  $\psi^{e}_{R}$ & &   & $-2$ & & & & 3 &   \\ [3pt]
  $\psi^{\nu}_{L}$ & &   & 0  & & & &   & 3 \\ [3pt]
  $\psi^{\nu}_{R}$ & &   & 0  & & & & 3 &   \\ [3pt]
\hline
  $\chi_L$ & 2 & & 1 &  &  & \\ [3pt]
  $\chi_R$ & & 2 & 1 &  &  & \\ [3pt]
  $Y_u$ &  &  &  & $ \bar{3}$ & 3 & \\ [3pt]
  $Y_d$ &  &  &  & $\bar{3}$ &  3  & \\ [3pt]
  $Y_\ell$ & & & & & & &$\bar{3}$ &  3\\[3pt]
$Y_\nu$ & & & & & & &$\bar{3}$ &  3\\[3pt]
\hline
\end{tabular} \end{center}
\caption{Model content. For ease of readability, horizontal lines separate
the quark multiplets, the lepton ones and the Higgs and flavon ones from
each other, and only non-singlet transformation properties are reported
explicitly.}
\label{tab:model}
\end{table}
\normalsize
One can clearly note the one-to-one correspondence between the
quark and the lepton multiplets, differing only in the behavior
under $SU(3)_c$. It is easy to verify that this field content
makes $G_{LR}$ completely anomaly-free, separately in the quark
and lepton sectors.

We next discuss the quark Yukawa couplings. We will ignore for the
moment the leptonic ones since they do not have any effect on the
final results for the quark sector. (The leptonic flavor
symmetries are discussed in sec.~\ref{sec:lepton}.) In writing the quark 
Lagrangian at the TeV scale, we will generally assume that the gauge 
symmetry $SU(2)_L\times SU(2)_R\times U(1)_{B-L}$ is restored at that
scale. Even under this assumption, one has still to specify where
the parity symmetry is broken. As anticipated in the Introduction,
one can either suppose that parity is restored at the TeV scale,
or else that its restoration takes place at some much higher scale
$M_P$ \cite{CMP}. Let us first focus on the former case, namely of
TeV-scale parity. In this case, the Lagrangian for the quark sector 
reads 
\bea 
\label{eq_Lquarks}
{\cal{L}}_{\rm q} &=& {\cal{L}}_{\rm q}^{\rm kin} - V(Y_u, Y_d,
\chi_L, \chi_R) + \lambda_u (\bar{Q}_L \tilde \chi_L \psi^{u}_{R}
+ \bar{Q}_R \tilde \chi_R \psi^{u}_{L}) + \lambda_d (\bar{Q}_L
\chi_L \psi^{d}_{R} +
\bar{Q}_R \chi_R \psi^{d}_{L}) \nn \\
&& + \lambda_u' \bar{\psi}^{u}_{L} Y_u \psi^{u}_{R} + \lambda_d'
\bar{\psi}^{d}_{L} Y_d \psi^{d}_{R} + {\rm h.c.}~,
\eea
where we have written explicitly only the Yukawa interactions.
We note at this point that, since under parity $Q_L\leftrightarrow Q_R$ 
and $\psi^{u}_{L} \leftrightarrow \psi^{u}_{R}$ (and similarly for 
$\psi^d_{L,R}$), parity symmetry requires 
$Y_{u,d} \leftrightarrow Y^\dagger_{u,d}$ and the $\lambda_{u,d}$ 
as well as $\lambda'_{u,d}$ couplings to be real.%
\footnote{In particular, concerning $\lambda_{u,d}$, one 
can note that there is one single such coupling for either of the 
up-type or down-type quark interactions with heavy fermions.
Hence, one can remove possible phases in $\lambda_{u,d}$ by 
absorbing them in the $\psi^u$ and $\psi^d$ fields, respectively.}
Parity will thus be broken only by the different vevs of $\chi_{L,R}$ 
(the tilde on these fields in eq.~(\ref{eq_Lquarks}) indicates 
$\tilde{\chi}=\tau_2\chi^*$ for both L and R).

In the case where parity is broken at a scale $M_P$ much higher 
than the TeV \cite{CMP}, the interactions in eq. (\ref{eq_Lquarks}),
obtained from each other by the parity operation defined in the previous
paragraph, have in principle different couplings. For example
\bea
\label{eq_noparityexample}
\lambda_u (\bar{Q}_L \tilde \chi_L \psi^{u}_{R}
+ \bar{Q}_R \tilde \chi_R \psi^{u}_{L}) ~\to ~
\lambda_{uL} \, \bar{Q}_L \tilde \chi_L \psi^{u}_{R}
+ \lambda_{uR} \, \bar{Q}_R \tilde \chi_R \psi^{u}_{L}~,
\eea
because of different RGE running beneath the scale $M_P$.
Hence, similarly as in eq. (\ref{eq_noparityexample}), in the case of 
no TeV-scale parity we will distinguish the left and right instances 
of each gauge, $\lambda$ and $\lambda'$ couplings by an L or R 
subscript.

Concerning the breaking of the gauge groups, the flavor gauge 
group $SU(3)_{Q_L} \times SU(3)_{Q_R}$ is broken spontaneously 
by the vevs of $Y_u$ and $Y_d$ while the group $SU(2)_L \times SU(2)_R$ 
by the vevs of the Higgs doublets, $\chi_{L,R}$, as already mentioned.
In particular, we adopt the following vev normalization
\bea &&\<\chi_L\> = \left(
\begin{array}{c}
 0 \\ v_L
\end{array}
\right)~,~~~~~
\<\chi_R\> =
\left(
\begin{array}{c}
 0 \\ v_R
\end{array}
\right)~,
\eea
while diagonal $Y$ vevs will be denoted henceforth as $\<\hat Y_{u,d}\>$.

\subsubsection*{Fermion masses}

From eq.\ (\ref{eq_Lquarks}) one can read off the up-type fermion mass Lagrangian to be
$\mc L_m ~=~ \ov {\mc U}_L M_u \mc U_{R}$, with $\mc U = {\rm column}\{u, \psi^{u}\}$, each of the
$u$ and $\psi^{u}$ fields carrying a generation index. The mass matrix reads
\bea
\label{eq_Mud}
M_u =
 \left(
\begin{array}{cc}
0
& \lambda_u v_L \id_{3 \times 3}   \vspace{3mm}
\\
 \lambda_u v_R \id_{3 \times 3}
& \lambda_u' \< \hat Y_u\>
\end{array}
\right)~,~~~~~
M_d =
 \left(
\begin{array}{cc}
0
& \lambda_d v_L \id_{3 \times 3}   \vspace{3mm}
\\
 \lambda_d v_R \id_{3 \times 3}
& \lambda_d' \< \hat Y_d\>
\end{array}
\right)~. 
\eea 
For the time being, we assume the parameters $\lambda_u v_L$ 
and $\lambda_u v_R$ to be much smaller than any of the 
$\lambda'_u \< \hat Y_u \>_i$. (With the subscript $i$ in 
$\< \hat Y_{u(d)} \>_i$ we shall henceforth label the diagonal 
entries of the flavon vev matrices.) Then, to leading order in an 
expansion in the parameters 
$\frac{\lambda_u v_{L(R)}}{\lambda'_u} \< \hat Y_u \>_{i}^{-1}$, 
and analogous for the down sector, the above mass matrices 
assume the following diagonal form 
\bea
\label{eq_Mud-diag} \hat M_u \simeq
 \left(
\begin{array}{cc}
\frac{\lambda_u^2 v_L v_R}{ \lambda_u'  \< \hat{Y}_u \> }
&0  \vspace{3mm}
\\
0
& \lambda_u' \< \hat Y_u\>
\end{array}
\right)~,~~~~~
\hat M_d \simeq
 \left(
\begin{array}{cc}
\frac{\lambda_d^2 v_L v_R}{ \lambda_d'  \< \hat{Y}_d \> }
&0  \vspace{3mm}
\\
0
& \lambda_d' \< \hat Y_d\>
\end{array}
\right)~. 
\eea 
From eq.\ (\ref{eq_Mud-diag}) it is evident that off-diagonalities 
in the light-quark Yukawa couplings are inherited from 
off-diagonalities in the flavon vevs $\<Y_{u,d}\>$.
We note first that, even below the $v_L$ scale, it is always
possible to have one of the flavon vevs in diagonal form through
an appropriate redefinition of the $\psi^{u}_{L,R}$ and
$\psi^{d}_{L,R}$ basis (see eq.\ (\ref{eq_Lquarks})). We choose
$Y_d$ to be that particular flavon multiplet. This amounts to
three parameters, fixed by the down-type quark masses. $Y_u$ 
will then be chosen to have a vev pattern of the form 
\bea
\label{eq_Yud_vevs} 
\< Y_u\> ~=~ V^\dagger_R \< \hat Y_u\> V_L~,
~~~~~\< Y_d\> ~=~ \< \hat Y_d\>~, 
\eea 
with $V_{L,R}$ unitary.
Note that, as already mentioned,  $V_L = V_R = V$ in eq.\
(\ref{eq_Yud_vevs}) follows from the $\< Y_u \>$ vev pattern being
hermitian and hence parity symmetric. This amounts to six real
parameters and three phases. Two of these phases can be absorbed
as relative phases of two up-type quark fields relative to the
third one. This gives the six real parameters and one phase to fit
up-type quark
masses and the CKM matrix.%
\footnote{We also note incidentally that, from the point of view of our discussion, the
$\lambda_{u,d}^{\prime}$ couplings can in principle be absorbed into the definition of
the $Y_{u,d}$ vevs respectively. This effectively leaves as free parameters only
$\lambda_{u,d}$, besides the scale of the vev $v_R$ and the $SU(2)_R$ coupling $g_R$,
making the model very economical. 
}

In the basis of eq.\ (\ref{eq_Yud_vevs}), and again in the temporary approximation
$v_R \ll \< \hat Y_{u,d} \>$, the Yukawa couplings of the SM (defined from the
interactions $\overline u_L y_u u_R$ and $\overline d_L y_d d_R$) read
\bea
y_u = \frac{\lambda_u^2 v_R}{ \lambda_u'} V_L^\dagger~ \< \hat{Y}_u \>^{-1} V_R~,
~~~~~ y_d = \frac{\lambda_d^2 v_R}{ \lambda_d'} \< \hat{Y}_d\>^{-1}~.
\label{eq_smyukawa}
\eea
One can now rotate the $u_{L,R}$ fields as
\bea
\label{eq_masseigenstates}
u_{L(R)} = V_{L(R)}^\dagger \hat u_{L(R)}~,
~~~~~d_{L(R)} = \hat d_{L(R)}~.
\eea
\newcommand{\Vckm}{V_{\rm CKM}}%
In the hatted basis, $y_u$ is diagonal and $V_L$ is moved to
the $\ov u_L \gamma^\mu W_\mu d_L$ interaction. Therefore, $V_L$
can be interpreted as the CKM matrix, $\Vckm$. As already stated, in our
case of left-right symmetry we have strictly $V_L = V_R = V$ at the
scale $v_R$. However, since we are interested only in $v_R$ values not
very far from the electroweak symmetry breaking scale $v_L$,
the radiative corrections to the above relation between left and right
CKM's are expected to be small. Therefore, we will henceforth generally
identify
\beq
\label{eq_CKM}
V_L ~=~ V_R ~=~ \Vckm~,
\eeq
with caveats to be commented upon more below in the analysis.
Since quark masses are given by $y_i v_L$, we can draw some
conclusions from the approximate relations (\ref{eq_smyukawa}):

\begin{itemize}

\item[(i)] In the limit of $v_R \ll \< \hat Y_{u,d} \>_i$ the elements
of the diagonal $\< \hat Y_{u,d} \>$ matrices follow an inverted hierarchy
with respect to the quark masses \cite{GRV, babu}.

\item[(ii)] For a given value of $v_R$ and of the $\lambda^{(\prime)}$ couplings,
eqs.\ (\ref{eq_smyukawa}) or the corresponding exact expressions in sec.
\ref{sec:fermionmixing} allow to univocally fix the $\< \hat Y_{u,d} \>$ entries.
Since the $Y_{u,d}$ vevs set also the mass scale for the flavor gauge bosons
(see below in this section for details), the inverted hierarchy
mentioned in item {(i)} implies a similar hierarchy in new flavor changing
neutral current effects: the lighter the generations, the more suppressed
the effects \cite{GRV}. This is arguably one of the most attractive features
of the model. We will return to this quantitatively in sec.\ \ref{sec:FCNC}.

\item[(iii)] In the exact parity case, the mass matrices $M_{u,d}$, 
see eq.\ (\ref{eq_Mud}), lead to arg det[$M_{u,d}$]=0, implying that the strong
CP parameter at the tree level vanishes. The one loop calculation
for a more general case of this type was carried out in Ref.\ \cite{babu1}.
Using this result, we conclude that the model solves the strong CP
problem without the axion.

\end{itemize}
Since the condition $v_R \ll \< \hat Y_{u,d} \>_i$ may in general not hold
for all flavors, we need to give for the quark masses a more exact relation than
eq.\ (\ref{eq_Mud-diag}).
In fact, the fermion mixing matrices (\ref{eq_Mud}) can be diagonalized exactly.
This will be discussed in sec.\ \ref{sec:fermionmixing}, along with its phenomenological
consequences.

\subsubsection*{Flavor gauge boson masses}

The masses of the $SU(3)_{Q_L} \times SU(3)_{Q_R}$ gauge bosons ${G_i}_{L,R}$ $(i=1,...,8)$ are
obtained from the kinetic terms of $Y_u$ and $Y_d$ in the Lagrangian,
$\Tr\left( | D^{\mu} Y_{u,d} |^2 \right)$, where the covariant derivatives are
\bea
D^{\mu} Y_{u,d} = \partial^{\mu} Y_{u,d} - i g_H  {G^{\mu}_R} Y_{u,d}  + i g_H Y_{u,d}  {G^{\mu}_L}~.
\eea
The relevant mass terms read
\bea
{\cal L} &=& \Tr \left( | g_H {G^{\mu}_R}  \<Y_u\>  - g_H \<Y_u\>  {G^{\mu}_L} |^2 \right) +
 \Tr \left( | g_H {G^{\mu}_R}  \<Y_d\> - g_H \<Y_d\>{G^{\mu}_L} |^2 \right) \nn \\
&=& \frac{1}{2} \mc G_k (M_V^2)_{kl} \mc G_l~,
 \label{eq_massG}
\eea where $\mc G_k \equiv \{ G^a_{L}, G^a_{R}\}$ is a vector
containing the 16 fields in $G_{L,R}^\mu = G_{L,R}^{\mu a}
\frac{\lambda^a}{2}$. The $\mc G_k$ are rotated by an orthogonal
matrix $O$ such that $\mc G_k = O_{kj} \hat{\mc G}_j$, where
$\hat{\mc G}_i$ are mass eigenstates. 

One interesting point to note is that all the flavor gauge boson masses are determined 
by basically only one $Y$ vev, namely the largest of the $Y_u$ vevs $\< \hat Y_{u} \>_1$.
In fact, for the lightest among the $M_{\mc G_i}$, $\<\hat Y_{u} \>_1$ is multiplied
by two powers of the Cabibbo angle $\theta_C$ (in the limit $\theta_C \to 0$,
one gets at least one massless $\mc G_i$) and $\<\hat Y_{u} \>_1 \times 
\theta_C^2$ is larger than the second-largest $Y$ vev contribution, $\<\hat Y_{u} \>_2$.

\section{Phenomenology} \label{sec:phenomenology}

In the subsections of this section we will discuss the various observables that are expected
to provide a constraint (or else the possibility of a signal) for the model. Since in some cases
-- starting from the model spectrum -- the model predictions vary in a wide range, we found it
useful to explore these predictions with a flat scan of the model parameters. In the case
where parity is assumed to be restored at the TeV scale, ranges have been chosen as follows:

\begin{enumerate}
 \item \label{en:gH} $g_H \in [0.3, 0.9]$ and $v_R$ such that $M_{W_R} \in [0.2, 5]$ TeV (the
 bound on $M_{W_R}$ from applicable constraints is taken into account afterwards).
 \item \label{en:lambda} $\lambda_{u} \in [0.96, 5]$, $\lambda_{d} \in [0.1, 5]$, see discussion
 below eq.\ (\ref{eq_m_condition}).
 \item \label{en:lambdap} Setting the couplings $\lambda'_{u,d} = 1$, one effectively absorbs
 them into the definition of the $Y_{u,d}$ vevs, respectively (the relevant combination entering
 fermion mixing is $\lambda'_{u,d} Y_{u,d}$, see eqs.\ (\ref{eq_Mud})). This assumption is however
 restrictive for the mass spectrum of the flavor gauge bosons, that depends on $\< \hat Y_{u,d} \> $,
 but not on $\lambda'_{u,d}$, see eq.\ (\ref{eq_massG}). Therefore, we have also scanned
 $\lambda'_{u,d} \in [0.1, 5]$.
 \item \label{en:gSU2} Finally, we have taken $g_L = g_R \simeq 0.65$ for the $SU(2)_{L,R}$
 couplings.
\end{enumerate}
In the other scenario where parity is not a good symmetry at the TeV scale, all the left vs. right couplings
can be chosen as different from each other. Concerning the $SU(2)_{L,R}$ couplings, in \cite{CMP} examples
have been given of scenarios where $g_R/g_L \sim 0.70$ for a UV complete theory which conserves parity.
Here we therefore limit ourselves to the reference choice $g_R = 0.7 \cdot g_L$,  Concerning the other
parameters:
\begin{itemize}
 \item The left and right instances of the $g_{H}$ and $\lambda'_{u,d}$ couplings have been scanned in
 the same ranges as specified in items \ref{en:gH} and \ref{en:lambdap} respectively.
 \item With regards to item \ref{en:lambda}, we have scanned $\lambda_{uL} \in [0.96, 5]$ and the rest of
 the $\lambda$ parameters in $[0.1, 5]$.
 \item We have further enforced that the left vs. right instances of each coupling do not differ from each
 other by more than a factor of 5.
\end{itemize}
Two concluding comments concern the reality of the $\lambda^{(\prime)}$ couplings and the hermiticity
of the Yukawa vevs, implicitly assumed in the above items. From the discussion below eq. (\ref{eq_Lquarks}), 
one can argue that, in the case where parity is not a good TeV-scale symmetry, non-negligible complex 
phases may be present in (some of) the $\lambda^{(\prime)}$ couplings. This may in turn have an impact 
on CP violating observables, which are, however, not the main concern in this paper, for the reasons 
mentioned in sec. \ref{sec:WRmass}. Finally, departures from hermiticity in the Yukawa vevs correspond 
(see discussion beneath eq. (\ref{eq_Yud_vevs})) to assuming sensible departures from eq.\ (\ref{eq_CKM}).
Throughout this paper we neglect such effects. Again, we will comment on this assumption in 
sec.\ \ref{sec:WRmass}.

\subsection{Tree-level FCNC effects} \label{sec:FCNC}

The flavor gauge bosons $G^{\mu a}_{L,R}$ couple to the currents $\mc J^{\mu a}_{H L,R} \equiv
g_{H} \overline Q_{L,R} \gamma^\mu \frac{\lambda^a}{2} Q_{L,R}$. Similarly as in Ref.\ \cite{GRV},
these interactions give rise to new, tree-level, contributions to the 4-fermion operators
\bea
\label{eq_Qi}
&&Q_1^{q_j q_i} ~=~ (\ov q_i^\alpha \gamma^\mu_L q_j^\alpha)(\ov q_i^\beta \gamma_{\mu,L} q_j^\beta)~,\nn \\
&&\tilde Q_1^{q_j q_i} ~=~ Q_1^{q_j q_i}|_{L \to R}~,\nn \\
&&Q_5^{q_j q_i} ~=~ (\ov q_i^\alpha P_L q_j^\beta)(\ov q_i^\beta P_R q_j^\alpha)~,
\eea
with Latin and Greek indices on the quark fields denoting flavor and respectively color, and
where $P_{L,R} \equiv (1 \mp \gamma_5)/2$. In the quark mass eigenstates basis, the Wilson
coefficients of the above operators read
\bea
\label{eq_Ci}
C_1^{q_j q_i} &=& - \frac{g_H^2}{8} (M^2_V)^{-1}_{a,b} (V_L^q \lambda^a V_L^{q \dagger})_{ij}
(V_L^q \lambda^b V_L^{q \dagger})_{ij}~,\nn \\
\tilde C_1^{q_j q_i} &=& - \frac{g_H^2}{8} (M^2_V)^{-1}_{8+a,8+b} (V_R^q \lambda^a V_R^{q \dagger})_{ij}
(V_R^q \lambda^b V_R^{q \dagger})_{ij}~,\nn \\
C_5^{q_j q_i} &=& \frac{g_H^2}{2} (M^2_V)^{-1}_{a,8+b} (V_L^q \lambda^a V_L^{q \dagger})_{ij}
(V_R^q \lambda^b V_R^{q \dagger})_{ij}~,
\eea
where $q$ can be $u$ or $d$, and a sum over $a$ and $b$ in the range $1,...,8$ is understood.
The matrices $V^{u,d}_{L,R}$ rotating the $u,d$ fields from the flavor to the mass eigenbasis should
be chosen as
\beq
V^{u}_{L,R} = V_{L,R}~,~~~~~V^{d}_{L,R} = \id~,
\eeq
compatibly with eq.\ (\ref{eq_masseigenstates}) and in the approximation of neglecting the mixing
between quarks and heavy fermion states.

\FIGURE[b]{
\includegraphics[width=0.49\textwidth]{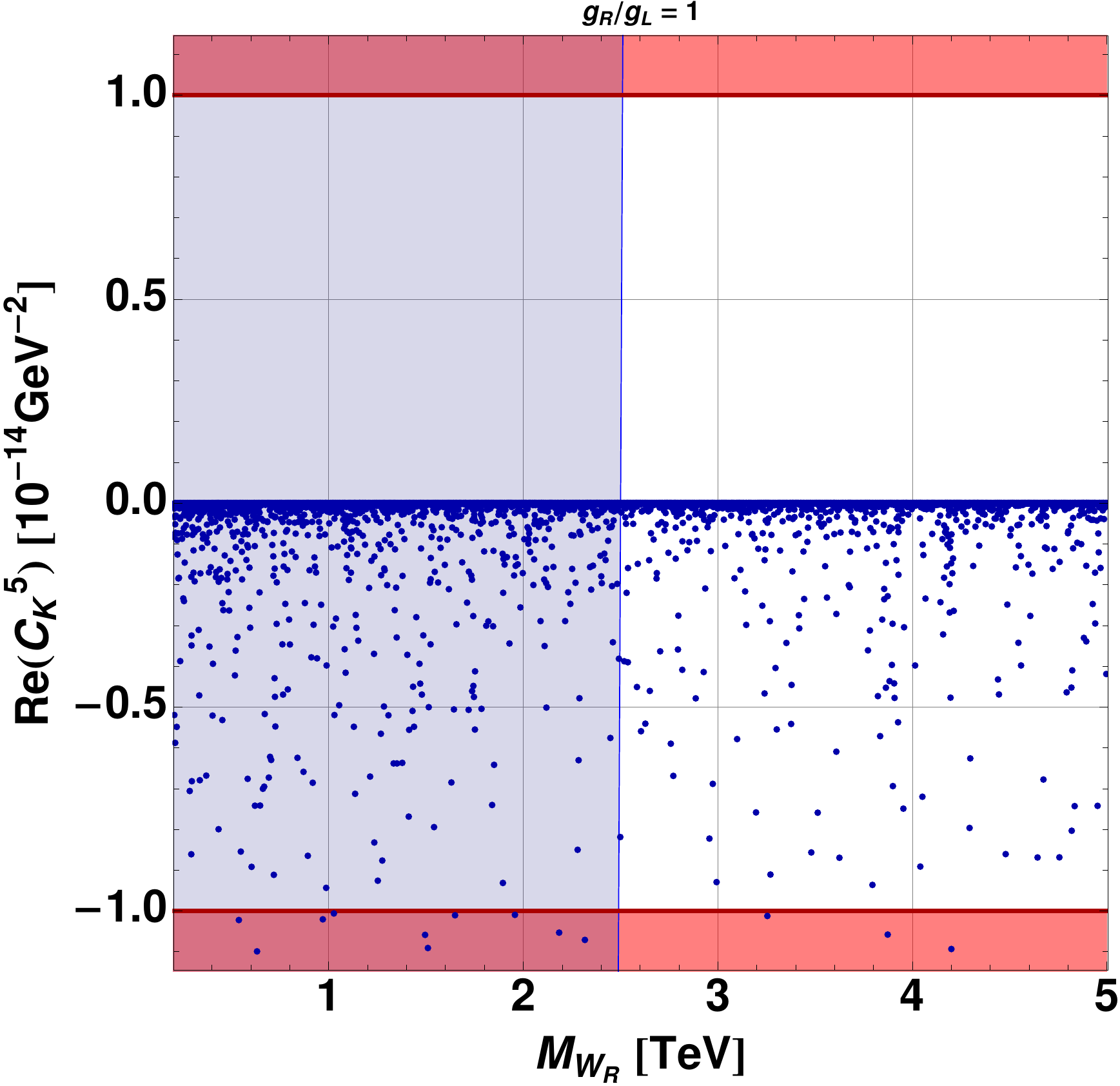}
\caption{$M_{W_R}$ vs. the contribution to the $\Delta F = 2$ Wilson coefficient $\re(C_K^5)$.
The horizontal (red) shaded regions are excluded from the analysis of Ref.\ \cite{UTfit_DF2}.
The vertical (blue) region extending leftwards is the $M_{W_R}$ bound from loop FCNCs \cite{cons}.}
\label{fig:bounds}
}
Updated bounds on the Wilson coefficients in eq.\ (\ref{eq_Ci}) have been reported by the UTfit
collaboration \cite{UTfit_DF2} and usefully tabulated in their table 4 for the different
meson-antimeson mixing processes. The contributions, predicted in our model, to the above coefficients
have been explored by the random scan mentioned at the beginning of sec.\ \ref{sec:phenomenology}.
As previously anticipated, these contributions are well within the existing bounds in the bulk of the
explored parameter space. As an illustration, we report in Fig.\ \ref{fig:bounds} the \textit{largest} in
magnitude among these contributions, that to the $K - \ov K$ mixing coefficient $\re(C_K^5)$,
in the case of TeV-scale parity.
The $M_{W_R}$ mass is therefore mostly bounded from box diagrams with $W_R$ exchange, as
discussed in sec.\ \ref{sec:WRmass}. The resulting bound, $M_{W_R} \gtrsim 2.5$ TeV \cite{cons},
is shown in Fig.\ \ref{fig:bounds} as a vertical shaded area extending leftwards.
On the $M_{G_i}$ masses, on the other hand, we will return in sec.\ \ref{sec:gaugemassscale}.

\subsection[Loop FCNC effects and lower bound on the $W_R$ mass]{Loop FCNC
effects and lower bound on the $W_R$ mass} \label{sec:WRmass}

\FIGURE[ht]{
\includegraphics[width=0.90\textwidth]{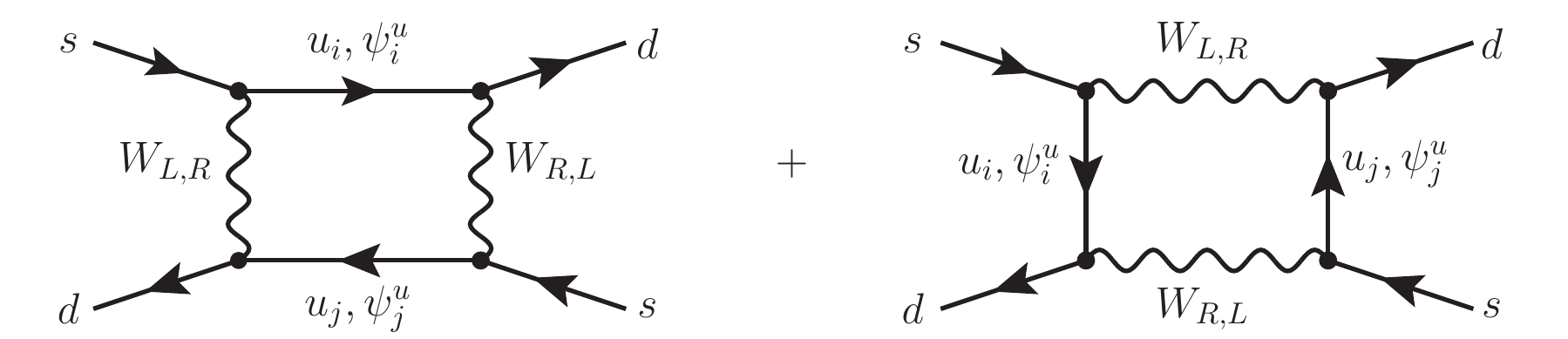}
\caption{Box diagrams contributing to the $K_L - K_S$ mass difference in our model.}
\label{fig:boxes}
}
In ref.\ \cite{cons}, it was pointed out that box diagrams with mixed $W_L$-quark
and $W_R$-quark exchange provide a severe lower bound on the $SU(2)_R$-symmetry
breaking scale in the case of the minimal LR symmetric model with Higgs bidoublets. This happens on account
of the constraint from the $K_L-K_S$ mass difference, $\Delta M_K$, and the bound reads
$M_{W_R} \gtrsim 2.5$ TeV.\footnote{Even more stringent bounds would come from CP violating observables
again in the $K$ sector, but these bounds are much more model dependent\cite{neme}.}

In our model, beside the known quarks, also their heavy fermionic
partners propagate in the box diagrams, because of fermion mixing.
The full amplitude, for a given quark-heavy fermion doublet (e.g.
$t$ and $\psi_{u,3}$), is depicted in Fig.\ \ref{fig:boxes}. The
sum of the contributions turns out to display a GIM-like mechanism
of cancellation, as we will shortly explain.

First note that the upper fermion line in any of the box diagrams of Fig.\ \ref{fig:boxes} will be proportional
to a factor of $V^\dagger_R \cos(\theta^u_{i,R}) m^u_{i} \cos(\theta^u_{i,L}) V_L$ or
$V^\dagger_R \sin(\theta^u_{i,R}) M^u_{i} \sin(\theta^u_{i,L}) V_L$ depending on whether the fermion is
a quark or its heavy partner (in the lower fermion lines one will have the hermitian conjugate of the same
expressions). Assuming propagation of a single quark species, it is therefore easy to see that
$\Delta M_K$ is such that
\beq
\label{eq_M12K}
\Delta M_K \propto g_L^2 g_R^2 \left | V_R^\dagger c_{i_R} c_{i_L} \left( \frac{m_i}{p^2 -m_i^2} -
\frac{t_{i_R} M_i t_{i_L}}{p^2 -M_i^2} \right) V_L \right |_{12}^2~,
\eeq
where we have abbreviated $\cos, \tan$ with $c, t$ and dropped the superscript on the masses.
One should note at this point that
\begin{itemize}
 \item[a.] the mass eigenvalues for quarks and heavy fermions come with opposite signs (cf. eq.
(\ref{eq_Mud})),
 \item[b.] $m_i$ and $t_{i_R} M_i t_{i_L}$ are equal, as can be seen by using eqs.\ (\ref{eq_m-m+}),
namely that the generally small mixing angles are compensated by a large mass in the second term in the
parenthesis of eq.\ (\ref{eq_M12K}),
 \item[c.] after factoring out the common mass term mentioned in item b, the diagram is proportional
to $m_i^2 - M_i^2$, as in the GIM mechanism. In our model the mass difference between a quark and its
fermionic partner is smallest in the top sector. Interestingly, in most of the meson mixings'
phenomenology of the down-sector, including the CP violating observable $\epsilon_K$, the top contribution
is the most important one.
\end{itemize}
The observation in item c has the potential of substantially weakening the severe bounds on the parity
breaking scale coming from $\epsilon_K$ \cite{cons}, and we reserve to come back to this issue in a separate
study. As stated elsewhere, the predictions for CP violating observables are however quite model-dependent,
and here we confine our discussion to CP conserving ones, in particular $\Delta M_K$. In this case,
the dominant loop contribution comes from the charm sector, hence the mechanism described above is much 
less effective, since the quark - heavy fermion splitting is very large. On account of this constraint, we find 
that the lower bound on the $W_R$ mass from $K_L-K_S$ mass difference coincides with that in ref. \cite{cons}, 
namely we get $M_{W_R} \gtrsim 2.5$ TeV. This bound holds in the exact parity case ($g_R/g_L = 1$),
that we have been assuming in this discussion.

On the other hand, if parity is broken at a scale much higher than the TeV scale, one
expects a splitting in the TeV-scale values of $g_L$ and $g_R$ in eq.\ (\ref{eq_M12K}), and values of
$(g_R/g_L)^2 < 1$ provide a further suppression of eq.\ (\ref{eq_M12K}) 
by the same factor. For example, assuming $g_R / g_L = 0.7$, this bound scales down to $M_{W_R} 
\gtrsim 1.7$ TeV. One can see this by simply noting that, as far as $M_{W_R}$ and the $SU(2)_{L,R}$
gauge couplings are concerned, $\Delta M_K$ scales as $\Delta M_K \propto g_L^2 g_R^2 / M_{W_R}^2$,
and that the $\Delta M_K$ calculation in the SM is dominated by loops mediated by the charm quark,
whose vectorlike partner $\psi^u_{2}$ is, to first approximation, decoupled.
As discussed at the beginning of sec. \ref{sec:phenomenology} the choice 
$g_R/g_L = 0.7$ \cite{CMP} will be our reference one for the scenario of no TeV-scale parity.

Two further comments are in order here. First, we note that a choice such as $g_R/g_L = 0.7$ will 
also affect the $W_R$ collider bound since the $W_R$ production rate will go down by the factor 
$(g_R / g_L)^2$ as well. Second, high-scale parity breaking will in general also cause some 
misalignment of the left and right CKM matrices.
In particular, large off-diagonal entries in the right CKM matrix have the potential of 
correspondingly increasing the contributions to flavor observables, a simple example being, again, 
that of meson anti-meson mixings.
The interest of this example is in the fact that the potential phenomenology of these contributions
encompasses not only flavor violation, but also {\em mixing}-induced CP violation, namely observables 
like $|\eps_K|$, $\sin 2 \beta$ and $\sin 2 \beta_s$.\footnote{On the other hand, we do not expect large
effects to these observables to come from the other potential sources of flavor mixing discussed in this 
paper, namely tree-level FCNCs mediated by flavor gauge bosons (see sec. \ref{sec:FCNC}) or quark 
-- heavy-fermion mixing (see sec. \ref{sec:bsgamma}).}

For the aims of the present discussion, we will assume that CKM entries undergo corrections due to RGE 
that don't modify their hierarchical structure, hence that the induced misalignment between $V_L$ and 
$V_R$ is small enough not to grossly alter the main argument of this section. A more detailed answer 
can be given, we feel, only in the context of specific models.

\subsection{Flavor gauge boson mass scale} \label{sec:gaugemassscale}

\newcommand{\G}{\mc G}
As mentioned above, from the point of view of flavor violating effects mediated by $\G_i$ exchange,
the model is compatible with $M_{\G_i}$ as small as O(TeV), and this represents a potentially interesting
new signal. Of course the question arises here, whether there are other model constraints placing a more
stringent lower bound on this mass.
\FIGURE[ht]{
\includegraphics[width=0.49\textwidth]{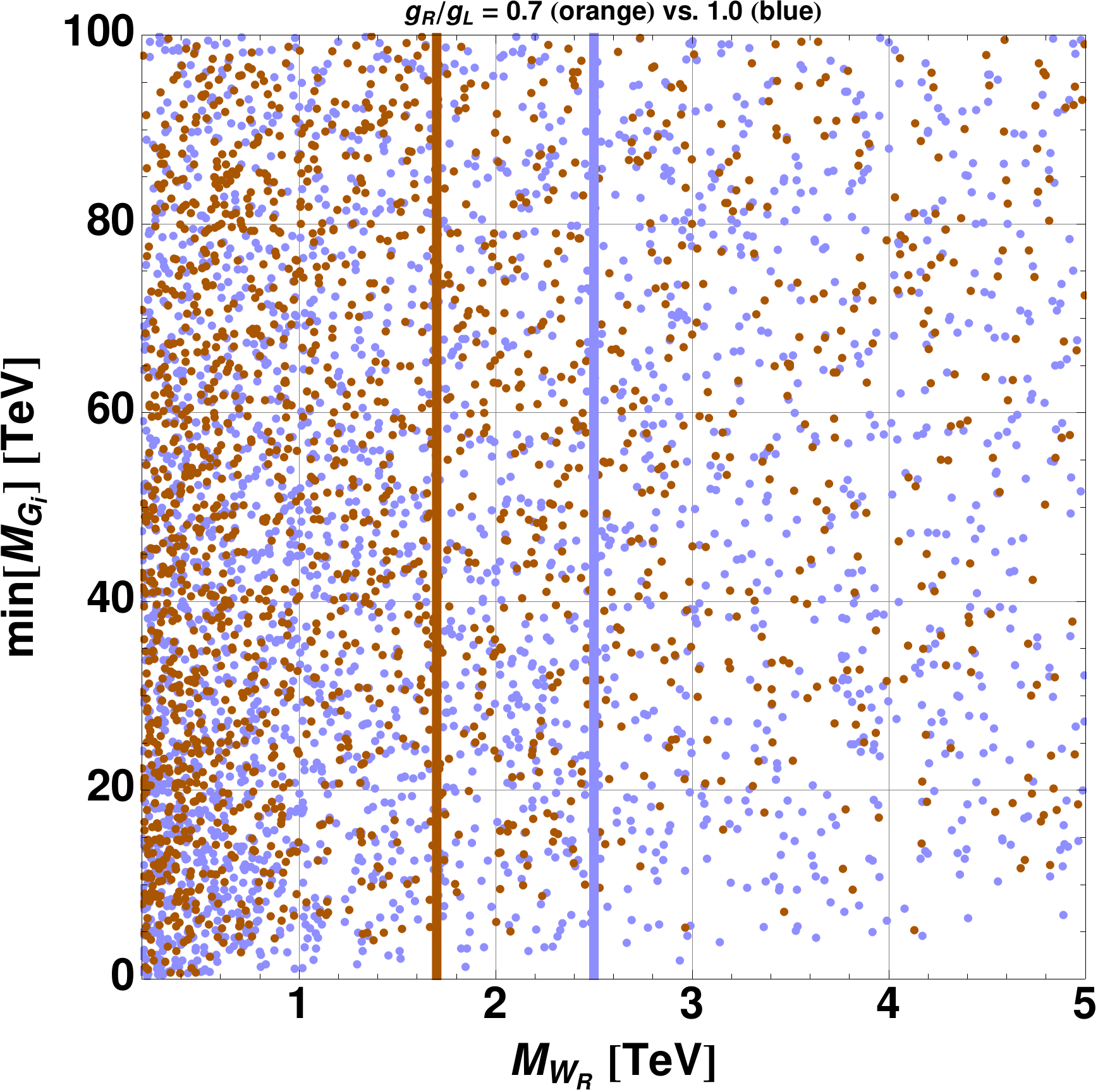}
\caption{$M_{W_R}$ vs. the mass of the lightest flavor gauge boson, min($M_{\G_i}$), for $g_R / g_L = 0.7$
(orange dots) or 1.0 (blue dots).}
\label{fig:MGi}
}
From eq.\ (\ref{eq_massG}), one expects the $\G_i$ spectrum to be correlated with the Yukawa vevs,
which are fixed, in the combinations $\lambda_u' \< Y_u \>$ and $\lambda_d' \< Y_d \>$, by the requirement
that, after fermion mixing, quark masses have the observed values. Therefore, with fixed $\lambda' \< Y\>$'s,
one may expect lower $\G_i$ masses to occur for larger $\lambda'$ couplings. This $M_{\G_i}$ vs. $\lambda'_{u,d}$
correlation is, however, completely smeared out by the freedom in the choice of the coupling $g_H$.

A correlation (albeit again not very sharp) is instead observed between the lowest allowed $\mc G_i$ mass 
and the $SU(2)_R$-breaking scale $v_R$, which, in turn, is related to the $M_{W_R}$ bound discussed in 
sec.\ \ref{sec:WRmass}. This correlation, and the applicable
$M_{W_R}$ bound, is shown in Fig.\ \ref{fig:MGi} in the two cases of exact TeV-scale parity (meaning $g_R / g_L = 1$)
and of no TeV-scale parity (where we assume, as mentioned $g_R / g_L = 0.7$ \cite{CMP}). From the lowermost
points, one can see that the minimum $M_{\G_i}$ mass tends to grow with growing $M_{W_R}$.
In particular, the lowermost red (blue) points on the right of the red (blue) vertical line imply allowed 
$M_{\G_i}$ values going down to about 3 TeV.

\subsection{Fermion mixing and its consequences} \label{sec:fermionmixing}

The fermion mixing matrix $M_u$ in eq.\ (\ref{eq_Mud}) can be diagonalized via orthogonal transformations
acting separately on each generation $i$ and each chirality, namely:
\beq
\label{eq_qQ_rotations}
\left(
\begin{array}{c}
 \hat u_i \\
 \hat \psi^u_i
\end{array}
\right)_{L(R)}
= \mc R(\theta^u_{L(R),i})
\left(
\begin{array}{c}
 u_i \\
 \psi^u_i
\end{array}
\right)_{L(R)}~,
~~~~~\mbox{with  }
\mc R(\theta) \equiv \left( \begin{array}{cc}
\cos \theta & - \sin \theta \\
\sin \theta & \cos \theta \\
\end{array} \right)~,
\eeq
where the hats denote mass eigenstates.
This eingenvalue problem admits the following analytic solution (we recall again that the diagonal entries
of the flavon vev matrices are denoted with $\< \hat Y_{u(d)} \>_i$)
\newcommand{\hYui}{\< \hat Y_{u} \>_i}
\newcommand{\hYu}{\< \hat Y_{u} \>}
\newcommand{\lahYui}{\lambda'_u \< \hat Y_{u} \>_i}
\bea
\label{eq_analytic}
&&(m^\pm_{u,i})^2~=~ \frac{1}{2} \left( (\lahYui)^2 + \lambda^2_u (v_L^2 + v_R^2) \pm \Delta^u_i
\right)~,\nn \\
&&\mbox{with   }\Delta^u_i ~\equiv~
\sqrt{(\lahYui)^4 + 2 (\lahYui)^2 \lambda^2_u(v_L^2 + v_R^2) + \lambda^4_u (v_L^2 - v_R^2)^2}
\eea
with rotation angles given by
\bea
\label{eq_angles}
&&\tan \te^u_{L,i} ~=~ \frac{\lambda_u^2 (v_L^2 - v_R^2) - (\lahYui)^2 + \Delta^u_i}{2 \lambda_u v_L \lahYui}~,\nn \\
&&\tan \te^u_{R,i} ~=~ \frac{\lambda_u^2 (v_R^2 - v_L^2) - (\lahYui)^2 + \Delta^u_i}{2 \lambda_u v_R \lahYui}~.
\eea
Provided the $\lambda, \lambda'$ couplings are of the same order, we will generally have
$(\lambda_u v_L)^2 \ll (\lambda_u v_R)^2, (\lahYui)^2$. Accordingly expanding the above
formulae one obtains the following approximate, but accurate solution ($x_i \equiv (\lambda_u v_L / \lahYui)^2$)
\bea
\label{eq_m-m+}
&&(m^-_{u,i})^2 ~=~ \frac{\lambda_u^4 v_L^2 v_R^2}{\lambda_u^2 v_R^2 + (\lahYui)^2} \,+\, {\rm O}(x_i^2)~,~~~~~
(m^+_{u,i})^2 ~=~ \lambda_u^2 v_R^2 + (\lahYui)^2 \,+\, {\rm O}(x_i)~,\nn \\
&&\tan \te^u_{L,i} ~=~ \frac{\lambda_u v_L \lahYui}{\lambda_u^2 v_R^2 + (\lahYui)^2} \,+\, {\rm O}(x_i^{3/2})~,~~~~~
\tan \te^u_{R,i} ~=~ \frac{\lambda_u v_R}{\lahYui} \,+\, {\rm O}(x_i)~.
\eea
Here $m^-$ and $m^+$ are to be identified respectively with the quark and heavy partners masses.
A completely analogous solution exists in the down sector and is obtained from the above
formulae with just the substitution $u \to d$. Note that the quark masses implied by the approximate $y_{u,d}$
in eq.\ (\ref{eq_smyukawa}) can be obtained from the $m^-_{u,i}$ in eq.\ (\ref{eq_m-m+}) in the limit
$\lambda_u v_R \ll \lambda_u' \hYui$.

Let us focus on the general solution in eqs.\ (\ref{eq_analytic}) and (\ref{eq_angles}). For fixed $\lambda_u$ and
$v_R$, one can determine the combination $\lahYui$ by inverting the equations $m^-_{u,i=1,2,3} = m_{u,c,t}$,
with the up-type quark mass values on the r.h.s. These equations admit a real solution in $\lahYui$ only for
sufficiently high $\lambda_u$. Note in fact that, with fixed values for the other parameters, the maximum of
$(m^-_{u,i})^2$ occurs for $\lahYui = 0$ (see also the first of eqs.\ (\ref{eq_m-m+})).
Necessary condition for a real solution to exist is therefore
\beq
\label{eq_m_condition}
m_{u,c,t} < m^-_{u,i}(\lahYui=0)|_{i=1,2,3} = \lambda_u v_L~,
\eeq
where we have assumed $v_L < v_R$. Since $v_L \simeq 174$ GeV -- very close to $m_t$ -- the above inequality
implies $\lambda_u \gtrsim 0.94$. (From the analogous inequalities in the down-quark sector, one also derives
$\lambda_d \gtrsim 0.02 $.) Note as well that, for the boundary value $\lambda_u = 0.94$, the solution of
eq.\ (\ref{eq_m_condition})
for $i = 3$ is $\lambda'_u \hYu_3 = 0$.
This explains the lower bound chosen
for $\lambda_u$ in our scans, see beginning of sec.\ \ref{sec:phenomenology}.

\subsubsection*{Parity-broken case}
In the case where parity is not a good symmetry at the TeV scale, we would expect the light quark Yukawa
couplings to be left-right asymmetric as already noted. In this case, the formulae for quark as well as
heavy-fermion masses and mixings change accordingly, i.e. eqs.\ (\ref{eq_m-m+}) are replaced by
\bea
\label{eq_m-m+P}
&&(m^-_{u,i})^2 ~=~
\frac{\lambda_{uL}^2\lambda_{uR}^2 v_L^2 v_R^2}{\lambda_{uR}^2 v_R^2 + (\lahYui)^2} \,
+\, {\rm O}(x_i^2)~,~~~~~
(m^+_{u,i})^2 ~=~ \lambda_{uR}^2 v_R^2 + (\lahYui)^2 \,+\, {\rm O}(x_i)~,\nn \\
&&\tan \te^u_{L,i} ~=~ \frac{\lambda_{uL} \, v_L \lahYui}{\lambda_{uR}^2 v_R^2 + (\lahYui)^2} \,
+\, {\rm O}(x_i^{3/2})~,~~~~~
\tan \te^u_{R,i} ~=~ \frac{\lambda_{uR} \, v_R}{\lahYui} \,+\, {\rm O}(x_i)~.
\eea
This case becomes very similar to the GRV examples \cite{GRV} and as we see from the figures below
allows vectorlike quark masses of about 2 TeV, making them, in principle, accessible at the LHC.
If $\lambda_{u(d) R}\ll \lambda_{u(d)L}$, then the vectorlike quark masses could be even lighter, as is
clear from eq.\ (\ref{eq_m-m+P}).

\FIGURE[t]{
\includegraphics[width=0.49\textwidth]{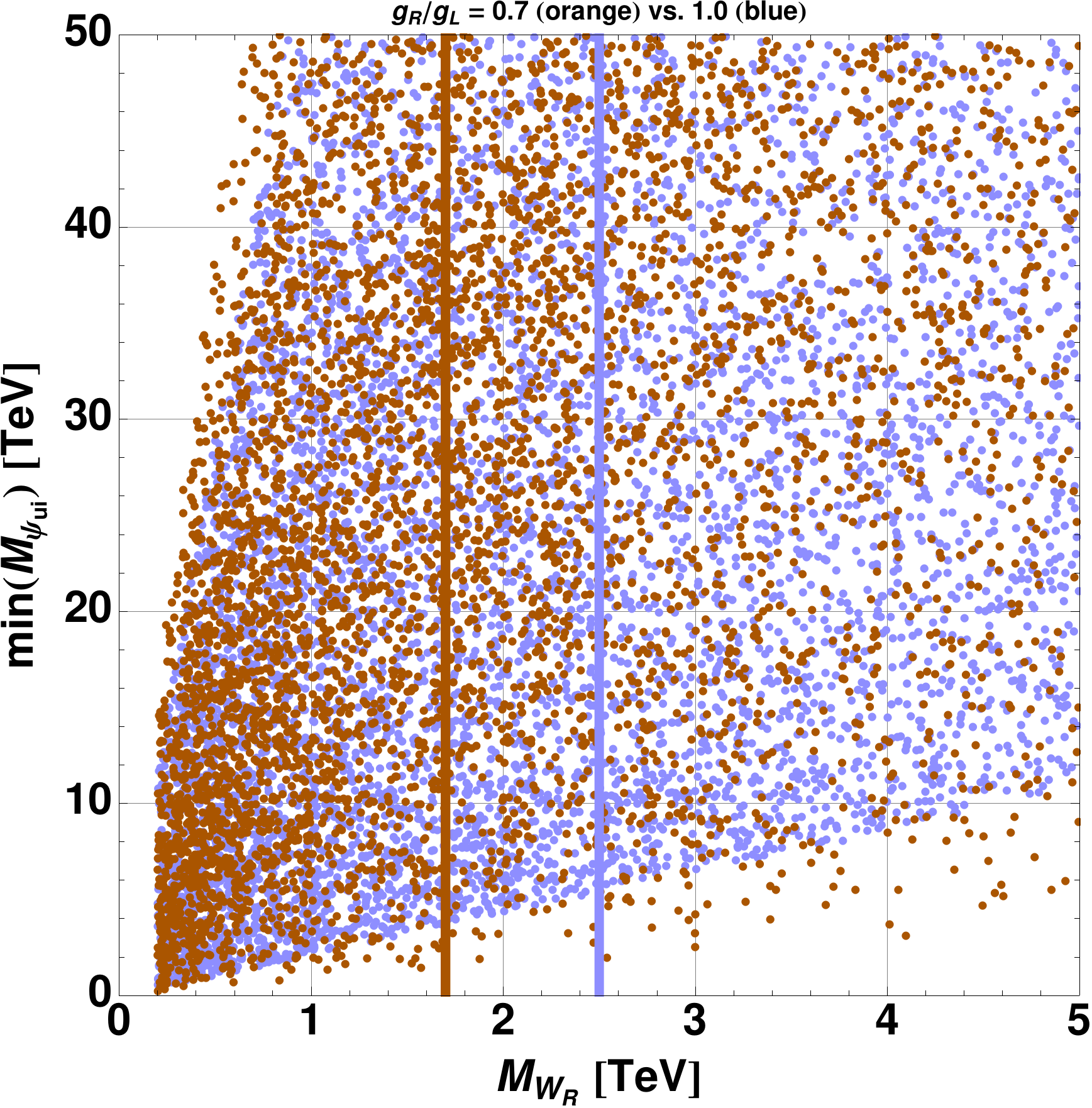}
\caption{$M_{W_R}$ vs. the mass of the lightest (up-type) fermionic partner, for $g_R / g_L = 0.7$
(orange dots) or 1.0 (blue dots). The vertical lines -- again in orange or blue for $g_R / g_L = 0.7$ or $1$ --
represent the $M_{W_R}$ bound discussed in sec.\ \ref{sec:WRmass}.}
\label{fig:minMpsi}
}
The one difference from the TeV-scale parity case is that there is no reason for $\lambda_{u,d}^\prime$ to
be real and therefore the model does not solve the strong CP problem. However, if there is parity
restoration at some high scale, at that scale one does have a solution to the strong CP problem and
an extrapolation is necessary to estimate how large a $\theta$ is induced at low energy. This kind of
analysis is beyond the scope of this paper and we hope to take it up separately.

\subsubsection*{Bounds on heavy-fermion masses and mixings}
The interesting phenomenological question is that of the magnitude
of the mixing angles and of the lowest allowed masses
$M_{\psi^u_i}$ and $M_{\psi^d_i}$ for the heavy up-type and
down-type fermion partners. Note that these masses are given by
the $m^+_{u,i}$ and $m^+_{d,i}$ solutions in eq.\ (\ref{eq_m-m+}).
According to these equations, the heavy fermion masses are
correlated with both the $SU(2)_R$-breaking scale $v_R$ and with
the scales of gauge flavor symmetry breaking. In practice the
latter correlation is blurred by the dependence on the unknown
$\lambda$ and $\lambda'$ parameters. On the other hand, the
correlation with $v_R$ still allows to infer the lowest allowed
values for $M_{\psi^u_i}$ and $M_{\psi^d_i}$, taking into account
the $M_{W_R}$ bound discussed in sec.\ \ref{sec:WRmass}. The
situation is illustrated in Fig.\ \ref{fig:minMpsi}, that displays
the lightest up-type fermion partner mass vs. $M_{W_R}$ . Blue and
respectively orange dots refer to the parity vs. no-parity scans,
see beginning of sec.\ \ref{sec:phenomenology} for details. The
vertical lines represent the corresponding $M_{W_R}$ bounds. One
can see that, while the exact parity case seems to exclude
$M_{\psi^u_i} \lesssim 5$ TeV, in the no parity case masses going
down to 2 TeV or even lower are possible. We mention that we found
similar values to be possible also for the lightest down-type
heavy fermion masses. In fact, as evident already from the first
of eqs.\ (\ref{eq_m-m+}), a large $m_t / m_b$ ratio does not
necessarily imply a corresponding hierarchy in $\lambda'_d \< \hat
Y_{d} \>_3 / \lambda'_u \< \hat Y_{u} \>_3$ because $\lambda_u^4 /
\lambda_d^4$ can be substantially larger than 1.

Further qualitative information can be obtained from eqs.\ (\ref{eq_angles}) for the mixing angles.
For $i = 1,2$, given the very large values of the $\hYui$, one can expect vanishingly small left-
and right-sector mixing angles.
For the top case, $i = 3$, the left-sector mixing angle is still generally small. In fact, note that the quantities
$\Delta^u_3$ and $\lambda_u^2 v_R^2 + (\lambda'_u \hYu_3)^2$ are very similar in size, and appear
with opposite signs in the numerator of $\tan \theta^u_{L,3}$. On the other hand, in the right-sector case, the
$\lambda_u^2 v_R^2$ term appears with reversed sign, and this, depending on the choice of parameters,
may result in a non-negligible $\theta^u_{R,3}$.

\FIGURE[ht]{
\includegraphics[width=0.48\textwidth]{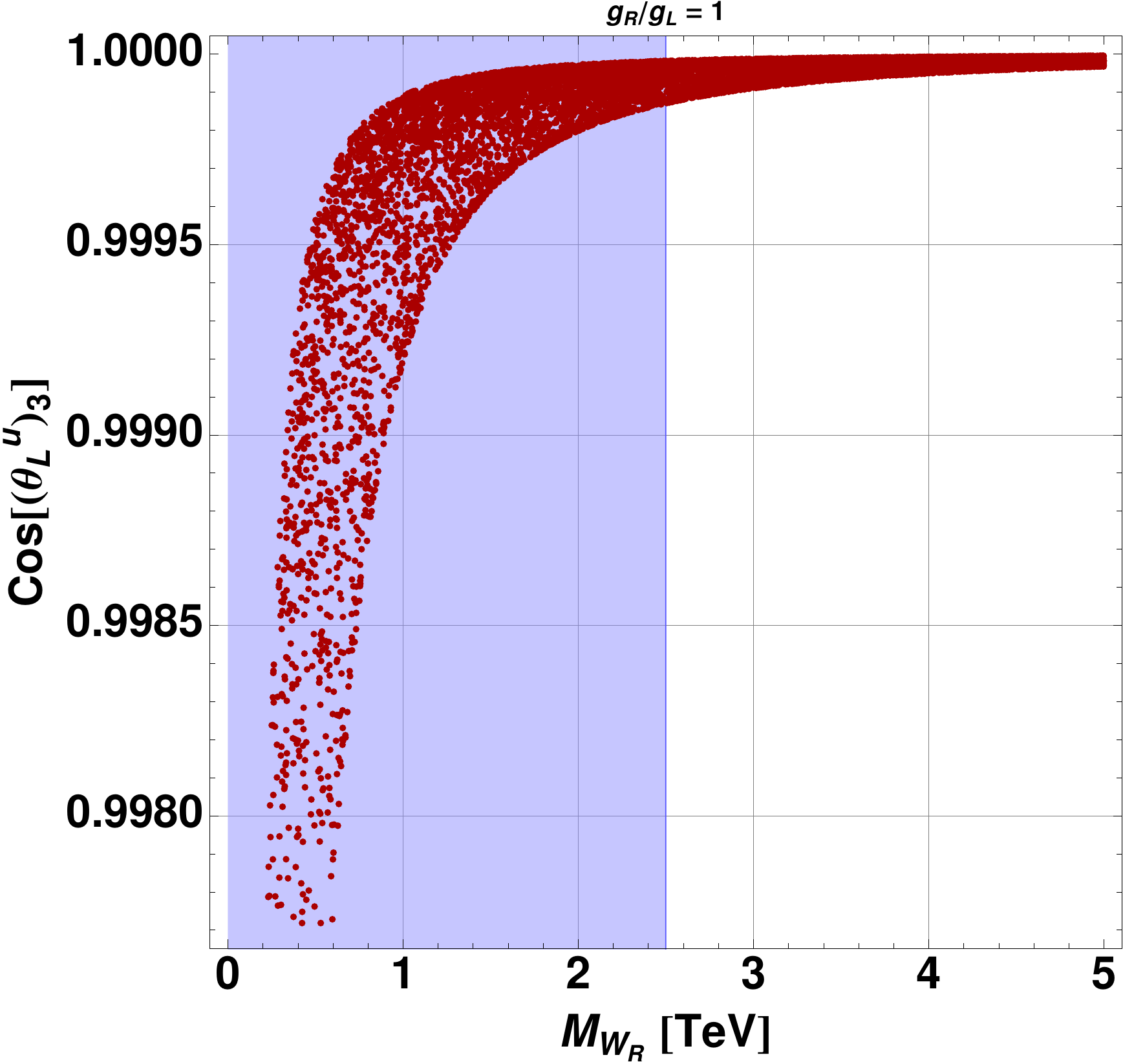}\hfill
\includegraphics[width=0.48\textwidth]{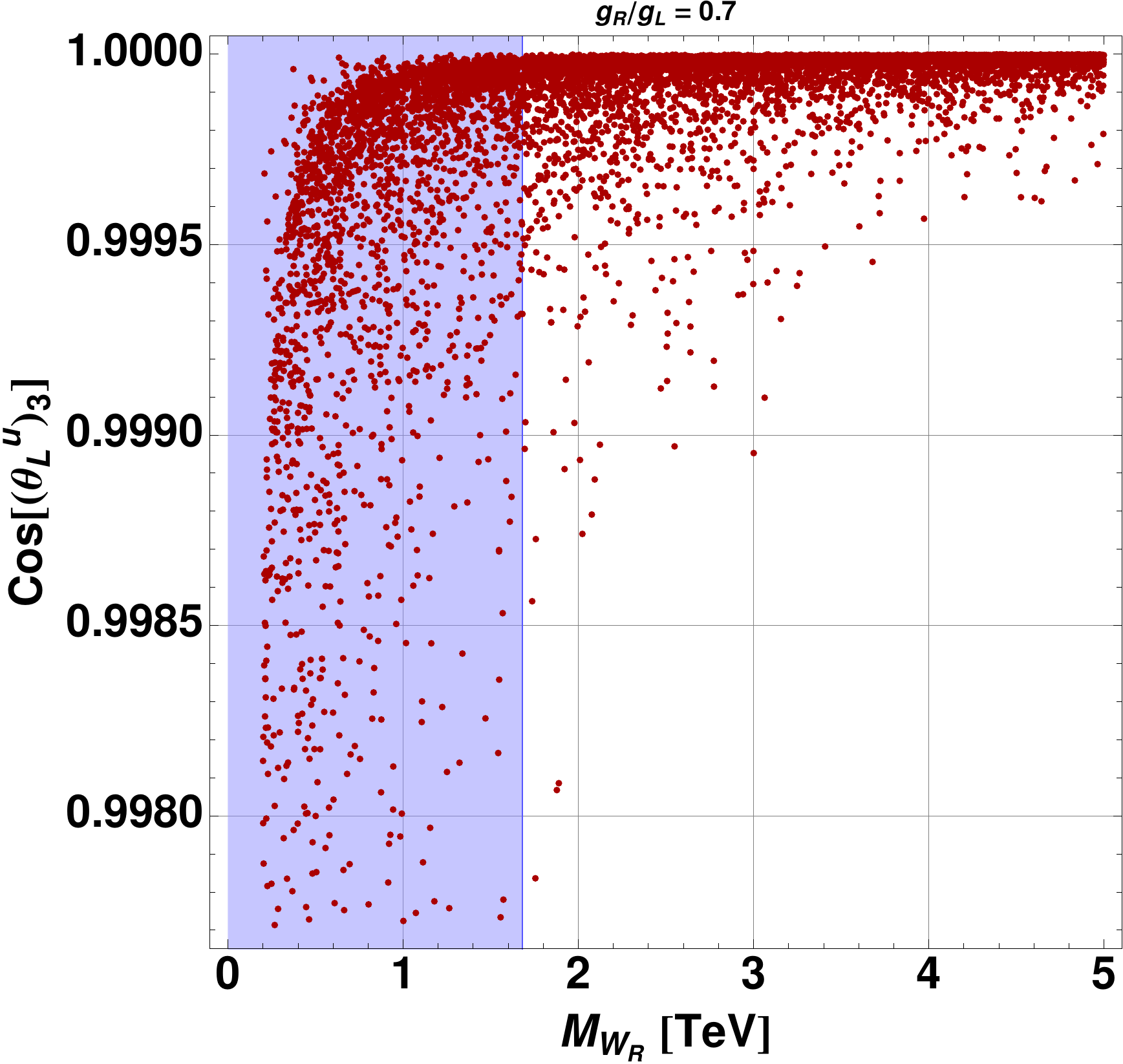}\vspace{0.2cm}
\includegraphics[width=0.48\textwidth]{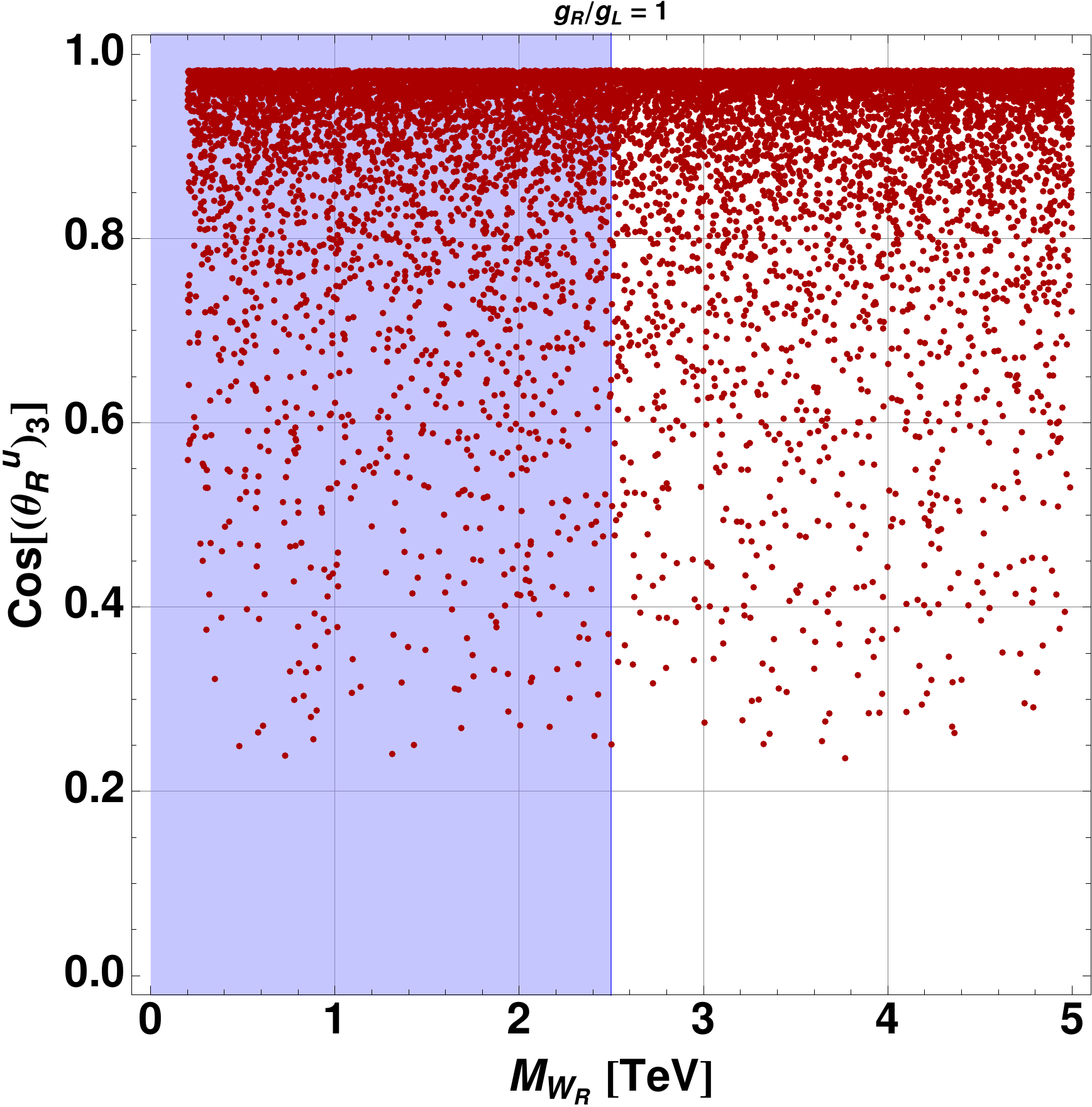} \hfill
\includegraphics[width=0.48\textwidth]{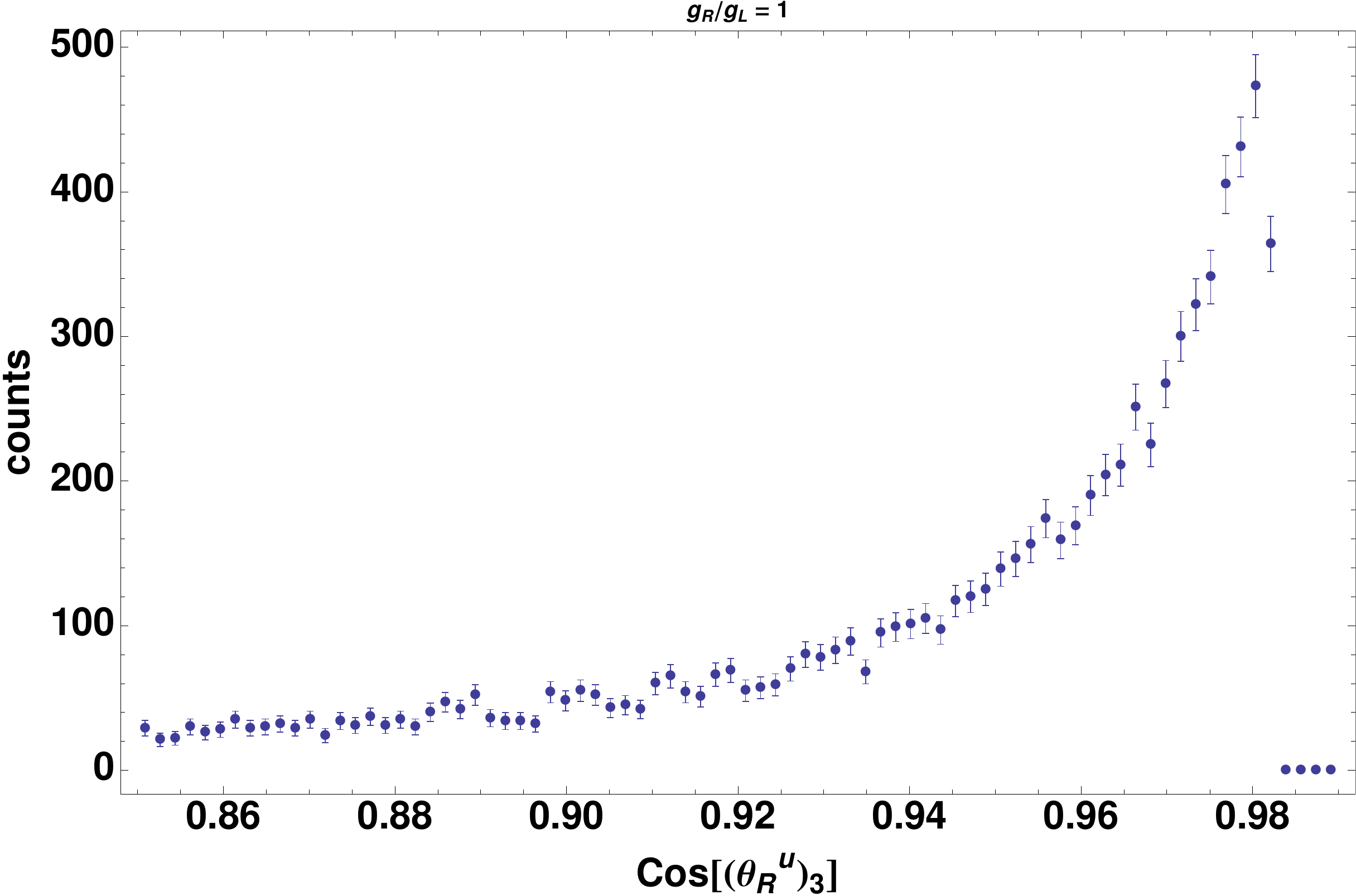}
\caption{{\em Upper panels:} $M_{W_R}$ vs. the fermion-mixing angles in the left-handed top sector, for
$g_R / g_L = 1$ {\em (leftmost panel)} or $0.7$ {\em (rightmost panel)}. {\em Lower panels:}
$M_{W_R}$ vs. the fermion-mixing angles in the right-handed top sector for $g_R / g_L = 1$, and
corresponding histogram (see text for further comments).}
\label{fig:fermionmixing_CosLR}
}
Fig.\ \ref{fig:fermionmixing_CosLR} illustrates the above considerations more quantitatively. The upper
panels show the mixing angles in the left-handed top sector against the $M_{W_R}$ mass, in the parity (left
panel) and in the no-parity case (right panel). One can see that mixing in the left-handed sector is always
fairly small -- taking into account the $M_{W_R}$ bound discussed in sec.\ \ref{sec:WRmass}, mixing is
such that $\sin (\theta^u_{L,3}) \lesssim 2 \times 10^{-2}$.
The corresponding angle in the right-handed sector is displayed in the lower panels for the case of TeV-scale
parity (the case of no parity gives similar results). In this case, the amount of mixing is not affected at all
by the $W_R$ mass bound, and we typically find $\theta^u_{R,3} \approx 10^\circ$ or larger, as also
displayed in the histogram. This implies $\sin (\theta^u_{R,3}) \gtrsim 0.2$, and may lead to potential
effects in observables like FCNC top decays such as $t \to c Z$ and observables sensitive to operators
with 4 powers of the top field.

In short, mixing angles anywhere else than in the $t_R$ case are vanishingly small. E.g., for $b_L$, we
find $\sin (\theta^d_{L,3}) \lesssim 2 \times 10^{-3}$. This bound is relevant to effects in observables
like $V_{tb}$ and $R_{b}$, on which we will be more quantitative in sec.\ \ref{sec:EWPT}.

\subsubsection{Electroweak precision tests} \label{sec:EWPT}

\subsubsection*{\boldmath $Z^0 \to b \bar b$}

The decay $Z^0 \to b \bar b$ is an example of the prototype process $V \to f \bar f$, with $V$ any of the
massive vectors, $f$ any of the fermions in the model, and $m_V \ge 2 m_f$. This kind of processes allows
to estimate the magnitude of tree-level non-oblique corrections that the model introduces.

The interactions relevant to $Z^0 \to b \bar b$ are those in $\mc L_{Zbb} \equiv Z^0_\mu J^\mu_{Z}$,
where $J^\mu_Z$ in our case is as follows
\beq
\label{eq_Zbb}
J^\mu_{Z} ~=~ \frac{g}{c_w} \left( \ov b_L \gamma^\mu (T_3^d c_{b_L}^2 - s_w^2 Q_d ) b_L + \ov b_R \gamma^\mu
(- s_w^2 Q_d ) b_R \right) + ...~,
\eeq
where $T_3^d$ and $Q_d$ are the eigenvalues of the weak isospin $\tau^3$ and electric charge operator of down-type quarks,
and dots denote the couplings to fermions other than the $b$. For ease of readability, we also abbreviate
$\cos \theta^{d}_{L(R),3} = c_{b_{L(R)}}$ and $\sin \theta^{d}_{L(R),3} = s_{b_{L(R)}}$. Note that the modification
with respect to the
SM current is only in the left-handed coupling proportional to $T_3^d$, because the electric charge operator is
diagonal across the quark and heavy-fermion fields, hence it commutes with the rotation (\ref{eq_qQ_rotations}).
The relevant Lagrangian (\ref{eq_Zbb}) is entirely analogous to that of Ref.\ \cite{GRV} hence the correction
to $\Gamma(Z^0 \to b \bar b)$ in our model will be the same, namely
\beq
\label{eq_dZbb}
\frac{\delta \Gamma(Z^0 \to b \bar b)}{\Gamma(Z^0 \to b \bar b)_{\rm SM}} ~=~
- s_{b_L}^2 \frac{2 + 4 s_w^2 Q_d}{1 + 4 s_w^2 Q_d + 8 s_w^4 Q_d^2} + O(s_{b_L}^4)~.
\eeq
To get a numerical idea of the correction implied by eq.\ (\ref{eq_dZbb}), one can first note that, since $s_w^2$ is a
small number, the coefficient of $-s_{b_L}^2$ is a number close to 2. Then one can recall, from our previous
numerical analysis of fermion mixing, that all the $s_{q_i}$ are tiny in the bulk of the model parameter space.
In particular, we quoted $s_{b_L}^2 \lesssim 4 \times 10^{-6}$ in the general discussion of fermion mixing.
Therefore, the constraint from $Z^0 \to b \bar b$ plays a completely
irrelevant role in our case, in comparison with direct searches of new heavy bosons (we will be back to this later on).
In fact, by establishing a lower bound on, e.g.,\ $M_{W_R}$, these searches constrain the fermion-mixing angles to
have values even closer to zero (see again Fig.\ \ref{fig:fermionmixing_CosLR}).

We note that, from the above argument, the most interesting effects of decays of the kind
$V \to f \bar f$ are expected in the top sector, in particular top production and decays.

\subsubsection*{Oblique corrections}

The $S, T$ and $U$ parameters~\cite{STU} quantify the modifications in the vacuum polarization diagrams for
the $SU(2)_L \times U(1)_Y$ vector bosons, due to the fact that the fermionic currents coupled to them are
altered with respect to their SM form. In our model, this occurs because of fermionic mixing and because
the new fermions have non-trivial charges under $U(1)_Y$. Following customary notation \cite{Peskin}, currents
are normalized as
\beq
J_a^\mu = \sum_f \bar{f} \gamma^\mu T^a f~,~~~~~~~J_Y^\mu = \sum_{f} \bar{f} \gamma^\mu Y_f f~,
\eeq
where $T^a$ denotes the $SU(2)_L$ representation, namely the identity or else $\sigma^a/2$, $Y_f$ the hypercharge
assignment for fermion $f$, and the sum runs over all fermions reported in the table of sec.\ \ref{sec:model} (our
fermion definition includes helicity projectors).

The $S$ and $T$ parameters are defined as \cite{STU}
\beq
\label{eq_ST}
S \equiv - 16 \pi \Pi'_{3Y}(q^2)|_{q^2=0}~,~~~~~~~
T \equiv \frac{4 \pi}{s_w^2 c_w^2 M_Z^2} ( \Pi_{11}(0) - \Pi_{33}(0))~,
\eeq
where $\Pi'$ denotes $d \Pi(q^2)/dq^2|_{q^2=0}$ as usual.
Their computation is a simple algebraic problem, after defining the `master' vacuum polarization amplitudes,
as the amplitudes with two left-handed currents or respectively one left- and one right-handed current at the two
vertices, and fermions of masses $m_1$, $m_2$ running in the loop.\footnote{Specifically, a left- or a right-handed
current means an insertion of $i \gamma^\mu \frac{(1 \mp \gamma^5)}{2}$ at the vertex, with namely no other overall
factor involved, e.g. color factors.} These amplitudes are denoted as $\Pi_{LL}(m_1^2,m_2^2,q^2)$ and
$\Pi_{LR}(m_1^2,m_2^2,q^2)$, and we shall follow the definition in \cite{Peskin}, that we do not rewrite here explicitly.

For the $T$ parameter we find
\beq
\label{eq_dT}
\delta T ~=~ \frac{3 \pi}{s_w^2 c_w^2 M_Z^2} \left( - 2 s_{t_L}^2 \, \Pi_{LL}^{bt}
+ s_{t_L}^2 (2 - s_{t_L}^2) \, \Pi_{LL}^{tt} + 2 s_{t_L}^2 \, \Pi_{LL}^{bT} -2
c_{t_L}^2 s_{t_L}^2 \, \Pi_{LL}^{tT} - s_{t_L}^4 \, \Pi_{LL}^{TT} \right)~,
\eeq
where $\delta T$ indicates that we have subtracted the pure SM contribution, obtained in the limit of no fermion mixing.
For ease of readability, we have also denoted $\Pi_{LL}(m_X,m_Y,0) = \Pi_{LL}^{XY}$, and, again, abbreviated
$\cos \theta^{u}_{L(R),3} = c_{t_{L(R)}}$, $\sin \theta^{u}_{L(R),3} = s_{t_{L(R)}}$.

For $\delta S$ we have instead
\bea
\label{eq_dS}
\delta S ~=~ 4 \pi \Bigl(- s_{t_L}^2 (2 - 3 s_{t_L}^2) \, \Pi_{LL}^{'\,tt} + 4 s_{t_L}^2 \, \Pi_{LR}^{'\,tt}
+ 6 c_{t_L}^2 s_{t_L}^2 \, \Pi_{LL}^{'\,tT} \phantom{\Bigl)~,}\nn \\
-(4 c_{t_L}^2 s_{t_L}^2 + s_{t_L}^4) \, \Pi_{LL}^{'\,TT} - 4 s_{t_L}^2 \, \Pi_{LR}^{'\,TT} \Bigl)~.
\eea

At this point, we note explicitly that eqs.\ (\ref{eq_dT}) and (\ref{eq_dS}) are obtained in the approximation of neglecting
fermion mixing other than in the top sector and (this is relevant only for $\delta S$) of including in the loops, among
the heavy fermions, only the top partner.
This is an excellent approximation, given the mass hierarchy among fermionic partners and the size of the mixing angles,
discussed before. Both of the $\delta T$ and $\delta S$ corrections turn out to depend only on
the LH mixing angle $\theta^{u}_{L,3}$ -- in $\delta T$ the dependence on $\theta^{u}_{R,3}$ combines in such a way
to cancel out in the final result. This fact is very welcome in our case, since, as discussed in sec.\ \ref{sec:fermionmixing},
$\theta^{u}_{R,3}$ is the only angle sizably different from zero. The allowed experimental values for eqs.\ (\ref{eq_dT})
and (\ref{eq_dS}) are of O$(10^{-2})$, with errors of O($10^{-1}$). Since $s_{t_L}^2$ is about $10^{-3}$, even
for $M_{W_R}$ as low as 500 GeV, the above corrections, similarly as $\delta \Gamma(Z^0 \to b \bar b)$, turn out to play
no constraining role at all.

\subsubsection[The decay $\ov B \to X_s \gamma$]{\boldmath The decay $\ov B \to X_s \gamma$}\label{sec:bsgamma}

Similarly as in \cite{GRV}, a further potential constraint for our model implied by flavor mixing comes from the
BR$(\ov B \to X_s \gamma)$, which is very accurately calculated within the SM \cite{Misiak} and also very precisely
measured experimentally \cite{HFAG10}. The two figures read, respectively (the photon energy cut is in both cases
$E_\gamma > 1.6~{\rm GeV}$.)
\bea
&&{\rm BR}(\ov B \to X_s \gamma)_{\rm exp} ~=~ (3.55 \pm 0.24 \pm 0.09)\times 10^{-4}~, \nn \\
&&{\rm BR}(\ov B \to X_s \gamma)_{\rm SM, NNLO} ~=~ (3.15 \pm 0.23) \times 10^{-4}~,
\eea
showing very good agreement with each other.

This decay in the SM is generated by a `magnetic-penguin' operator
induced by a $W-t_L$ loop. Its Wilson coefficient at the $W$
scale, $C_7(m_t, M_W)$, is modified in our model because the $t_L$
is not a mass eigenstate: $t_L = c_{t_L} \hat t_L + s_{t_L} \hat \psi^t_L$
(we have again used the shortcut $\cos(\theta^u_{L,3}) = c_{t_L}$). Neglecting the
running between the $\psi^t$ mass, here indicated as $m_T$
($\gtrsim 500$ GeV, as we discussed in sec.\ \ref{sec:fermionmixing}) and
the $W$ scale, this effect can be
accounted for by a shift in $C_7(m_t, M_W)$, \beq \label{eq_C7}
C_7(m_t, M_W) \to c_L^2 C_7(m_t, M_W) + s_L^2 C_7(m_T, M_W)~, \eeq
plus an analogous shift in the coefficient $C_8(m_t, M_W)$ of the
chromomagnetic penguin operator. Since $C_7(\mu \approx m_b)$
enters as $|C_7|^2$ in the branching ratio, the leading effect is
due to interference, and is of O($s_{t_L}^2$). To get a numerical idea
of the effect, one may use the next-to-leading order~(NLO) SM formulae of
\cite{Chetyrkin}. Including the shift (\ref{eq_C7}) and using $M_T
= 500$ GeV, we obtain
\beq
\label{eq_bsgammashift}
{\rm BR}(\ov B \to X_s \gamma) ~=~ (3.2 + 1.3 \, s_{t_L}^2) \times 10^{-4}~.
\eeq
In view of the smallness of $s_{t_L}^2$ in the bulk of our parameter space,
the above shift is well within the theoretical error.

\subsection{Further constraints} \label{sec:further}

\subsubsection{Electric dipole moments}

After diagonalizing the quark -- heavy-fermion mass matrix, all CP violating
fermion couplings arise from the $Y_u$ or $Y_d$ vevs. In particular, with
our choice of basis in eq.\ (\ref{eq_Yud_vevs}), they must be proportional to
$\< Y_u \>$. One may expect that one-loop diagrams with intermediate gauge
bosons (either $W_R$ or the flavor bosons $\mc G_i$) and up quarks, and one
quark mass insertion, may result in new contributions to the up quark EDM.
In the flavor-boson case, using eq.\ (\ref{eq_Mud-diag}), it is however easy
to convince oneself that the contribution to the EDM must be of the form
\beq
\label{eq_deu}
d_e^u \propto \frac{\lambda_u^2 v_L v_R}{\lambda'} \im \left(
\frac{\lambda^a}{2} V^\dagger_{\rm CKM} \< \hat Y_u \>^{-1} V_{\rm CKM}
\frac{\lambda^a}{2} \right)_{11}~,
\eeq
with $\lambda^a$ the Gell-Mann matrices. Similarly as the one-loop SM contribution,
the contribution in eq.\ (\ref{eq_deu}) vanishes trivially because of the
hermiticity of the matrix on the r.h.s.. A completely similar argument
holds of course in the $W_R$ case. Hence new contributions to quark EDMs
may arise in our model only at the two-loop level and are therefore very small.

\subsubsection{Top quark flavor changing effects}

Among the model predictions testable at the LHC are top-quark flavor changing effects,
e.g. a modification in the $\bar t c G$ coupling. In our model, neutral Higgs
interactions do not give rise to any flavor changing effect due to the fact that they
are diagonal.
However, the flavor gauge boson couplings involve the CKM matrix
as well as the flavor generators, both of which can mix
generations. We will do a detailed study of these effects in a
subsequent paper. Here we simply give an estimate of the dominant
contribution to the operator $\bar t \sigma_{\mu\nu} c G^{\mu\nu}$
to be of order
 \bea g_{tcG}~\sim ~ \frac{v_Lv_R}{16\pi^2(\<\hat{Y}_{u}\>_3)^3} \, ,
 \eea
which can be estimated to be of order 
$10^{-3}\left(\frac{\rm TeV}{\< \hat Y_u \>_3}\right)^3$ TeV$^{-1}$.
Such effects have been looked for at the Tevatron and will be looked for
in processes such as $GG\to t\bar c$, $cG\to t\gamma$, etc. at the LHC \cite{pedro}.
The current Tevatron (D{\O}) bound on the strength of such operators is
$\leq 0.018$ TeV$^{-1}$ with a 2.2 fb$^{-1}$ dataset \cite{tcg}.

\subsubsection{Direct searches} \label{sec:direct}

A key feature of models of this kind is the existence of three
heavy vectorlike families, which essentially helps to ameliorate
the severe FCNC bounds expected on the basis of dimensional
analysis. In this section we address the bounds on their masses
based on direct collider searches. The CDF collaboration has searched
for up-type heavy quarks (called generically $t'$ in the
literature) and provides a lower bound on their mass of 335 GeV \cite{tprime}.
Likewise, there is a lower limit on down-type heavy quarks, also from CDF,
giving $m_{\psi_d} \geq 385$ GeV \cite{bprime}.
These analyses assume the heavy quarks to decay 100\% of the time to a $W$
and light quarks. This will hold in our model for the lightest of the
vectorlike quarks.

\section{Lepton Sector} \label{sec:lepton}

Within our framework, the discussion of the lepton sector is completely
parallel to the quark sector as far as the flavor gauge boson and charged
lepton spectra are concerned. The relevant flavor gauge group is in this
case $SU(3)_{\ell_L}\times SU(3)_{\ell_R}$, and one introduces
two further flavon fields $Y_{\nu,\ell}$, transforming as $(\bar{3}, 3)$
under this group.  The gauge invariant Yukawa interaction for the leptons
is then completely analogous to eq.\ (\ref{eq_Lquarks}), but for the replacement
of quark doublets with lepton ones and heavy quark partners with heavy lepton
partners. Of course, the $\lambda$ and $\lambda'$ couplings also do not
need to be the same as those appearing in eq.\ (\ref{eq_Lquarks}).
The fermion mixing argument leading to eq.\ (\ref{eq_smyukawa}) is likewise
trivially generalizable to this case, hence for the diagonal elements
of $\<Y_\ell\>$ one expects the relation $ \< \hat{Y}_e \> :
\< \hat{Y}_\mu \>: \< \hat{Y}_\tau \> ~=~ m^{-1}_e
: m^{-1}_\mu : m^{-1}_\tau$.

\subsection{Neutrino masses}

Concerning the neutrino sector, after symmetry breaking the mass matrix 
for $(\nu_{L,R}, \psi^{\nu}_{L,R})$ separates into two block matrices 
involving $(\nu_L, \psi^{\nu}_{R})$ or $(\nu_R, \psi^{\nu}_{L})$.
For the first case we have 
\bea 
M_{\nu-N}~=~\left(\begin{array}{cc}0 &
\lambda_\nu v_L\\ \lambda_\nu v_R &
\<Y_\nu\>\end{array}\right)~,
\eea
and similarly for the $(\nu_R, \psi^{\nu}_{L})$, after exchanging 
$L\leftrightarrow R$ in the above matrix. As a result, we have two 
sets of Dirac neutrinos: $\nu_L$ pairing with $\psi^{\nu}_{R}$ and 
$\nu_R$ with $\psi^{\nu}_{L}$. In the limit of $\<Y_\nu\>\gg v_R$, 
the neutrino mass formula reads
\bea
\label{eq_Mnu}
M_\nu~=~\frac{\lambda^2_\nu v_L v_R}{\<Y_\nu\>}~.
\eea 
It is clear from the above equation that, if $v_R$ and $\<Y_\nu\>$ 
are in the few TeV range, we need to choose $\lambda_\nu \sim 10^{-6}$ 
in order to get the right order of magnitude for neutrino masses (in
the sub-eV range). Note that already this is an improvement over
the SM, where getting Dirac masses of the right order requires the
Yukawa coupling to be much smaller (of order $10^{-12}$).
Furthermore, we need to choose $\<Y_\nu\>$ in such a way as to get
the observed large neutrino mixings. As far as the $ \psi^{\nu}_{L,R}$
fields are concerned they will have masses of order of the flavor
symmetry breaking scale $\< Y_\nu \>$.

\subsection{Constraints}

The above setup is subject to various constraints.
First, since neutrinos are Dirac fermions, the right-handed neutrinos
have $W_R$-mediated interactions, that can keep them
in equilibrium with charged leptons, unless the right-handed
interactions are sufficiently weak. Therefore, the model will predict $N_\nu=6$
at the Big Bang Nucleosynthesis epoch, which is not consistent
with our current understanding of Helium, Deuterium and Lithium
abundances of the universe\cite{Nakamura:2010zzi}. In fact, this
leads to a lower bound on the mass of the right-handed $W_R$'s of
order $3.3$ TeV\cite{bbn}. It must however be noted that, if one
generates Majorana masses for the $\psi^{\nu}_{R}$ by adding
$SU(3)_R$ sextet Higgs fields with vev, one can lift the
right-handed neutrinos to higher masses and understand the
lightness of left-handed neutrino masses via the seesaw mechanism.
In this case, there is no lower bound on the $W_R$ mass from Big
Bang Nucleosynthesis.

Second, it is interesting to note a recent lower bound on the 
$W_R$ mass of 1.36 TeV from the CMS experiment at the LHC \cite{CMS}.
This bound directly applies to our model, and in general to models with 
Dirac neutrinos.

Third, within this generalization of the model, possible constraints 
on the flavor gauge boson scale may come from lepton-flavor violating 
(LFV) decays such as $\mu \to e \gamma$ and $\mu \to 3e$. The existing
limits on these decays can actually be used to estimate a lower bound
on the leptonic flavor scale as follows. Formula (\ref{eq_Mnu}) for 
the Dirac neutrino mass matrix can be trivially inverted to give
\bea
\label{eq_Ynu}
\< Y_\nu \> ~=~\frac{\lambda^2_\nu v_L v_R}{M_\nu}~.
\eea
This expression can then be rewritten using the formula
$M_\nu~=~U^*\hat{M}_\nu U^\dagger$ (where $\hat{M}_\nu$ is the
diagonal neutrino mass matrix) as
\bea
\< Y_{\nu} \>_{\alpha \beta}~=~\lambda^2_\nu v_Lv_R\sum_i
U^*_{\alpha i}U^*_{\beta i}m^{-1}_i~,
\eea
with $m_i$ the diagonal entries of $\hat{M}_\nu$.
The form of $\< Y_\nu \>$ clearly depends on the neutrino mass
ordering. Taking for simplicity normal ordering, $m_1 \ll
m_2 \ll m_3$, we get the dominant contribution to be
\bea
\< Y_{\nu} \>_{\alpha \beta}~=~\lambda^2_\nu v_Lv_R U^*_{\alpha 1}
U^*_{\beta 1}m^{-1}_1~.
\eea
To have an estimate of the typical $\< Y_\nu \>$ size, one may
choose $m_1 \sim 0.5 m_{\odot} \sim 0.005$ eV and use the
tri-bi-maximal form for the lepton mixing matrix $U$. For a TeV
$v_R$, we find $\< Y_\nu \>$ entries of $\sim$ 100 TeV. 

We can now provide an estimate of the decay rates for the processes 
$\mu \to 3e$ and $\mu \to e \gamma$. The amplitude for the $\mu \to 3e$ 
process arises at the tree level due to flavor diagonal and off-diagonal 
gauge boson mixing, namely from the terms $\<Y_{\nu}\>_{11}$ and 
$\<Y_{\nu}\>_{12}$. 
Since neutrino mixings are large, we assume these terms to be of 
similar size, indicated as $\overline Y_\nu$. Hence the amplitude 
has the form
\bea
\label{eq_ALFV}
A(\mu\to 3e) \sim \frac{1}{2 \overline Y_\nu^2}~.
\eea
This relation is nothing but a simplified version of eqs.\ ({\ref{eq_Ci}}). 
Note, in particular, that the gauge coupling dependence is of course absent, 
because the flavor-gauge boson masses also scale with it. To translate eq.
(\ref{eq_ALFV}) into a branching ratio, one can use the fact that the 
calculation of $\mu \to 3e$ is very similar to the well-known calculation 
of $\Gamma(\mu \to e \nu_\mu \ov \nu_e) \simeq \Gamma_{\mu, {\rm tot}}$,
but for the replacement of $G_F/\sqrt2$ with the amplitude in eq. 
(\ref{eq_ALFV}). Hence the branching ratio for $\mu \to 3e$ can
be simply estimated as 
\beq
\label{eq_BRmu3e}
B(\mu \to 3e) \sim \frac{1}{2 \overline Y_\nu^4 G^2_F}~.
\eeq 
This can be of order $10^{-12}$, like the current experimental limit 
\cite{mu3e}, for $\overline Y_\nu \sim 300$ TeV. Because of our assumption
of roughly equal entries in the $Y_\nu$ vev matrix, the same estimate 
applies to all the other LFV decays into three charged leptons, such
as $\tau \to 3e$ or $\tau \to 3\mu$.
The limits on these decays are (currently) much weaker \cite{LFV-tau3l} 
and as such satisfied for the above mentioned value of the leptonic flavor 
scale. The situation may of course change drastically in the event of new
data from a super flavor factory.

Turning to the $\mu \to e \gamma$ decay, it is generated by a loop graph,
and its branching ratio can be estimated to be of order
$\frac{27\alpha}{16\pi G^2_F \overline Y_\nu^4}$. Given the loop
suppression with respect to eq. (\ref{eq_BRmu3e}), one gets values safely 
below the current experimental limits \cite{muegamma} for the above choice 
$\overline Y_\nu \sim 300$ TeV.

A final comment is in order. The above discussion about LFV observables 
was mostly aimed at verifying that reasonable values for the relevant 
massive parameters of the model do not lead to conflicts with the current
LFV bounds. A separate and potentially interesting question not addressed 
in this paper is whether our setup may explain a positive LFV signal from 
current or planned experiments. While our arguments, in particular the one 
following eq. (\ref{eq_BRmu3e}), suggest a positive answer, a more detailed 
one requires invoking a specific flavor model to be embedded within our 
framework.

\section{Conclusions}\label{sec:conclusions}

We have examined the possibility of gauged flavor symmetry as a way to explore 
the origin of quark lepton masses and mixings. As was noted in Ref.\ \cite{GRV}, 
in such models there is an inverse correlation between the quark masses and the
flavor hierarchy between the gauge boson masses, making it possible to have 
light enough flavor gauge bosons and enhanced FCNC effects for the third generation.
We have worked within the left-right symmetric electroweak group, which
seems to provide a number of advantages over the SM gauge group
while maintaining this inverse relation. These advantages include a reduction in 
the number of input parameters, a possible solution to the strong CP problem without 
the axion (provided parity is also a TeV-scale symmetry), and the possibility of 
accommodating neutrino masses.
For the case where parity is a TeV-scale symmetry, the lower bounds on both the 
lightest vectorlike fermion mass as well as on the flavor gauge symmetry scale is of
about 5 and respectively 10 TeV (see Figs. \ref{fig:minMpsi} and \ref{fig:MGi}).
On the other hand, if only $SU(2)_R$, but not parity, survives as a good symmetry 
down to the TeV scale, the lightest phenomenologically allowed vectorlike quark mass 
could be much lower. The lightest flavor gauge boson mass gets likewise lower.
How low one can go down for these masses depends on what one assumes for the 
difference between the left and the right couplings, which in turn depends on the 
nature of the UV complete parity-symmetric theory.
We have noted the consistency of the model with all the best-known phenomenology, 
including electroweak precision data. The detailed predictions for the FCNC effects 
in the third generation case are currently under investigation; here we only made some 
qualitative comments about top flavor changing effects.

\acknowledgments
DG acknowledges useful discussions with Roberto Contino, Adam
Falkowski, Slava Rychkov and Giovanni Villadoro. The authors also
acknowledge Giovanni Villadoro for various comments on the manuscript.
The work of DG was supported by the EU Marie Curie IEF Grant no.
PIEF-GA-2009-251871. The work of RNM was supported by
the NSF grant PHY-0968854, and the work of IS was supported by
the U.S.~Department of Energy through grant DE-FG02-93ER-40762.

\bibliographystyle{JHEP}
\bibliography{LRwHS}

\end{document}